\documentclass[reprint,
superscriptaddress,
 amsmath,amssymb,
 aps,
pra,
]{revtex4-2}

\usepackage{graphicx}
\usepackage{dcolumn}
\usepackage{bm}
\usepackage{braket}
\usepackage{hyperref}
\usepackage{cancel}
\usepackage{float}
\usepackage[normalem]{ulem}
\hypersetup{colorlinks = true, allcolors = blue}

\begin{document}

\title{Energy-Resolved Eigenmode Spectroscopy of 1-D and 2-D Non-Hermitian Skin Effects}

\author{Rohith Srikanth}%
\affiliation{%
 Department of Electrical Engineering, University of Maryland, College Park, Maryland 20742, USA
}%

\author{Sashank Kaushik Sridhar}
 \affiliation{Department of Mechanical Engineering, University of Maryland, College Park, Maryland 20742, USA}

\author{Avik Dutt}
\email{avikdutt@umd.edu}
 \affiliation{Department of Mechanical Engineering, University of Maryland, College Park, Maryland 20742, USA}

\affiliation{
 Institute for Physical Science and Technology, University of Maryland, College Park, Maryland 20742, USA
}
\affiliation{
 National Quantum Laboratory (QLab) at Maryland, College Park, Maryland 20740, USA
}

\date{\today}
\begin{abstract}
Non-Hermitian lattices can host the non-Hermitian skin effect, a boundary-induced collapse of all bulk eigenstates into exponentially localized edge modes. This effect underlies anomalous bulk-boundary correspondence and remarkable enhancements in non-Hermitian sensing, yet direct energy-resolved access to the eigenmodes of non-Hermitian lattices has remained limited. Here we report band- and energy-resolved eigenmode spectroscopy of skin modes in a frequency synthetic dimension. By introducing strong frequency-domain boundaries in an electro-optically modulated ring resonator, we realize finite non-Hermitian lattices and use laser detuning as a spectroscopic axis for the eigenenergies of the effective Hamiltonian. Site-resolved heterodyne measurements then reconstruct the spatial profile of each mode, revealing boundary-localized skin states throughout the spectrum and their eigenenergy-dependent displacement from the edge. Beyond 1D, the same frequency-boundary architecture, upon incorporating long-range couplings between finite lattices, produces genuine 2D frequency lattices rather than the hitherto-realized folded 1D systems on twisted tubes. In these lattices we observe tunable directional transport and edge localization in two synthetic dimensions. Our results introduce eigenmode spectroscopy as a direct probe of non-Hermitian physics and establish strongly bounded frequency lattices as a flexible platform for Hamiltonian engineering.

\end{abstract}

\maketitle

Non-Hermitian (NH) topological models are distinguished by strong exponential localization of a macroscopic fraction of eigenstates at the boundaries, a phenomenon known as the non-Hermitian skin effect (NHSE) \cite{hatano_localization_1996,martinez_alvarez_non-hermitian_2018, yao_edge_2018}. 
In addition to this boundary localization feature, these systems exhibit unique physical behaviors that have garnered considerable attention in the context of sensing applications \cite{okuma_topological_2020, mcdonald_exponentially-enhanced_2020,Liu_non-hermitian_2021,Yuan_non-hermitian_2023,Parto_enhanced_2025,wang_observation_2025}. 
As a result, these models have been explored across a variety of platforms, including acoustic/mechanical systems \cite{ghatak_observation_2020,Zhang_observation_2021,Gu_transient_2022}, electronics \cite{Helbig_generalized_2020,Liu_non-hermitian_2021,zhu_higher_2023, Yuan_non-hermitian_2023}, cold atoms \cite{Liang_dynamic_2022,Zhao_two-dimensional_2025}, and photonics \cite{weidemann_topological_2020, Xiao_non-hermitian_2020,wang_generating_2021, wang_topological_2021}.  Among these, frequency-domain photonic platforms offer a uniquely scalable route, treating the resonant frequency modes of one or two coupled ring resonators as `synthetic' lattice sites and engineering inter-site coupling via intra-cavity electro-optic modulation \cite{wang_generating_2021,wang_topological_2021,Ye_observing_2025, orsel_giant_2025,blanchard_exponentially_2025,yu_comprehensive_2025}. However, these systems have predominantly been characterized through reciprocal-space ($k$-space) measurements, with little information about the lattice space. 

Remarkable signatures exist in both lattice-space occupations and in reciprocal-space band structures of topological systems, more so with non-Hermiticity in the mix. For example, NHSE in lattice-space is intrinsically linked to integer-quantized winding, point-gap topology, and line-gap topology in $k$-space, with the generalized bulk-boundary correspondence underpinning their physical connections ~\cite{koch_bulk-boundary_2020, Helbig_generalized_2020}. Hence, it is important to jointly measure signatures of NH behavior in both $k$-space and with bounded lattice-site occupation. Frequency synthetic dimensions present a unique advantage to achieve this, offering reconfigurable coupling geometries, flexible lattice sizes, strong boundaries, and high-contrast band structure measurements. Despite recent advances showing clear manifestations of non-Hermiticity in frequency synthetic dimensions  \cite{blanchard_exponentially_2025}, they are limited to a band-agnostic lattice occupation. Much richer physics than the now-routine across-the-band NHSE is revealed by combing through the full spectrum of eigenmodes that emerge at each possible eigenenergy \cite{Wang_non-hermitian_2024}. Such a direct band-energy-resolved characterization has remained elusive across all NH experiments, with only indirect inference of eigenmodes and eigenenergies in most systems \cite{Helbig_generalized_2020}.

In this work, we achieve lattice and band-resolved eigenmode spectroscopy in a frequency synthetic dimension platform with strong boundaries. Leveraging low-noise site-resolved readout enabled by a cascaded heterodyne scheme \cite{sridhar_measuring_2025}, we report one of the first measurements of the NHSE in frequency synthetic lattices and, to our knowledge, the first demonstration of energy-resolved eigenmode spectroscopy among non-Hermitian platforms. A key advance enabling these demonstrations is the realization of a stable, strong boundary in the synthetic-dimension system.
The same strong boundary also allows the realization of ``genuine'' 2-D lattices through long-range couplings, circumventing the issue of twisted boundary conditions that have been ubiquitous in previous frequency implementations. 
To do so, we measure the spread of an excitation in 2-D space, with independently tunable transport directionality along each axis. These results showcase the power and simplicity of frequency-domain synthetic photonic platforms for analog Hamiltonian simulation and mark an important step toward more versatile and robust quantum simulators and non-Hermitian sensors. 
\begin{figure*}[!htbp]
    \centering
    \includegraphics[width = \textwidth]{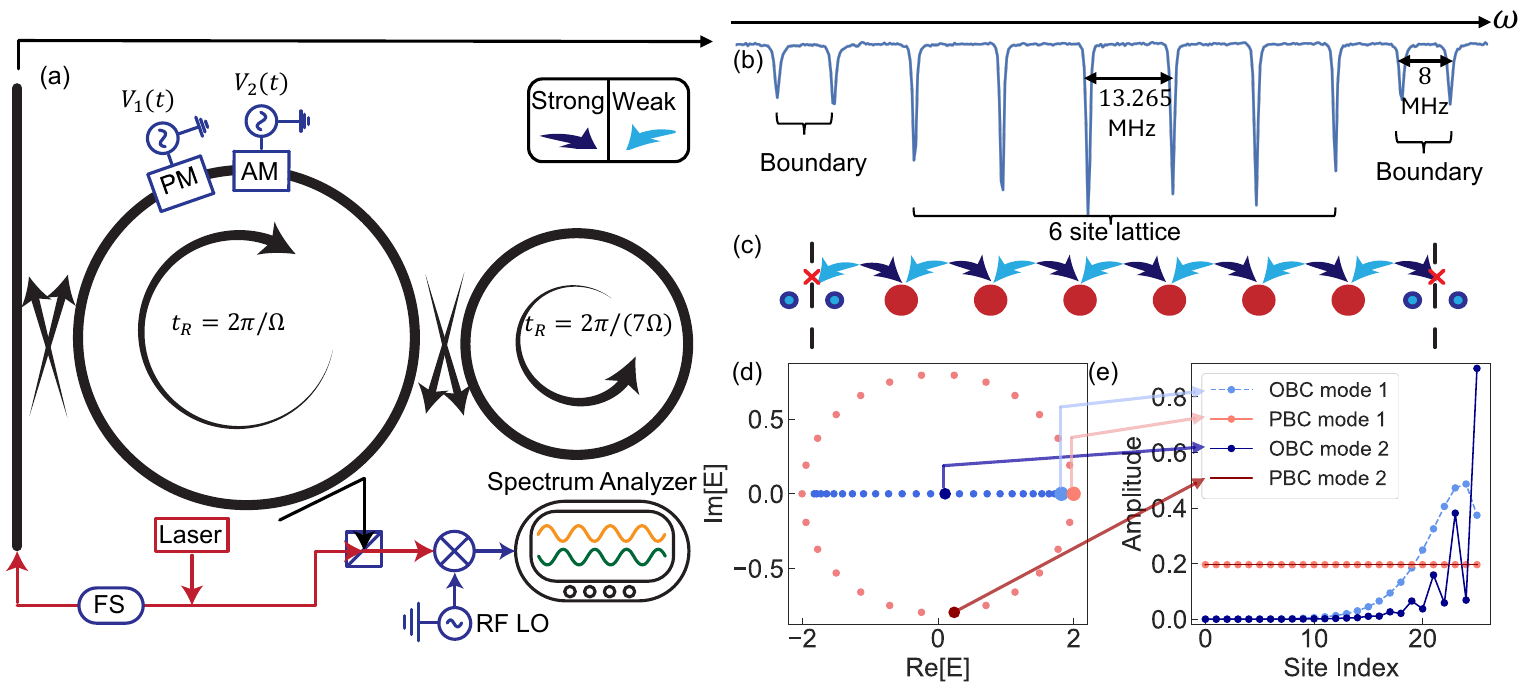}
    \caption{(a) Experimental schematic of a ring resonator that supports equally spaced frequency modes coupled to an auxiliary ring that is $7\times$ smaller in length, thus hybridizing every 7$^\text{th}$ resonance of the primary ring. The corresponding transmission of the system is measured experimentally and shown in (b). To construct 1-D lattices, the modes are coupled via electro-optic modulation through the phase modulator (PM) in the primary ring by driving them at a frequency corresponding to the lattice spacing. At the location of the split modes there is a local change in spacing preventing efficient coupling beyond it, effectively creating a frequency boundary at the mode splitting as depicted in (c). By including amplitude modulation (AM), we can implement 1-D non-Hermitian models that require nonreciprocal couplings between modes. (d) Complex energy spectrum of a 1-D non-Hermitian lattice under periodic boundary conditions (PBC - red) and open boundary conditions (OBC- blue). Two energies are highlighted for each set of boundary conditions - real part of the energy close to 0 and close to its maximum value. (e) Eigen mode amplitudes of the lattice corresponding to the energies highlighted in (d) .
    }
    \label{fig:Schematic}
\end{figure*}

One of the foundational 1-D non-Hermitian models is the Hatano-Nelson (HN) lattice, with nonreciprocal coupling strengths: $C+\Delta e^{i\varphi}$ and $C-\Delta e^{-i\varphi}$ in the forward and backward directions respectively. $C$ represents the Hermitian hopping strength, $\Delta$ represents the anti-Hermitian hopping strength and $\varphi$ describes the phase difference between the two. The extent of nonreciprocity can be controlled by tuning the relative strength and phase between these two coupling terms. The tight-binding description for the HN model is as follows:
\begin{gather}
H_{HN}=\sum_n (C+\Delta e^{i\varphi})c_{n+1}^\dagger c_{n}+(C-\Delta e^{-i\varphi})c_{n}^\dagger c_{n+1}
\end{gather}
where \(c_{m}^\dagger\) (\(c_{m}\)) represents the bosonic creation (annihilation) operators at the site indexed by \({m}\) where \(m\in \mathbb{Z}\) represents the lattice site number. (Fig.~\ref{fig:Schematic}). 

\begin{figure*}[ht!]
    \centering
    \includegraphics[width = \textwidth]{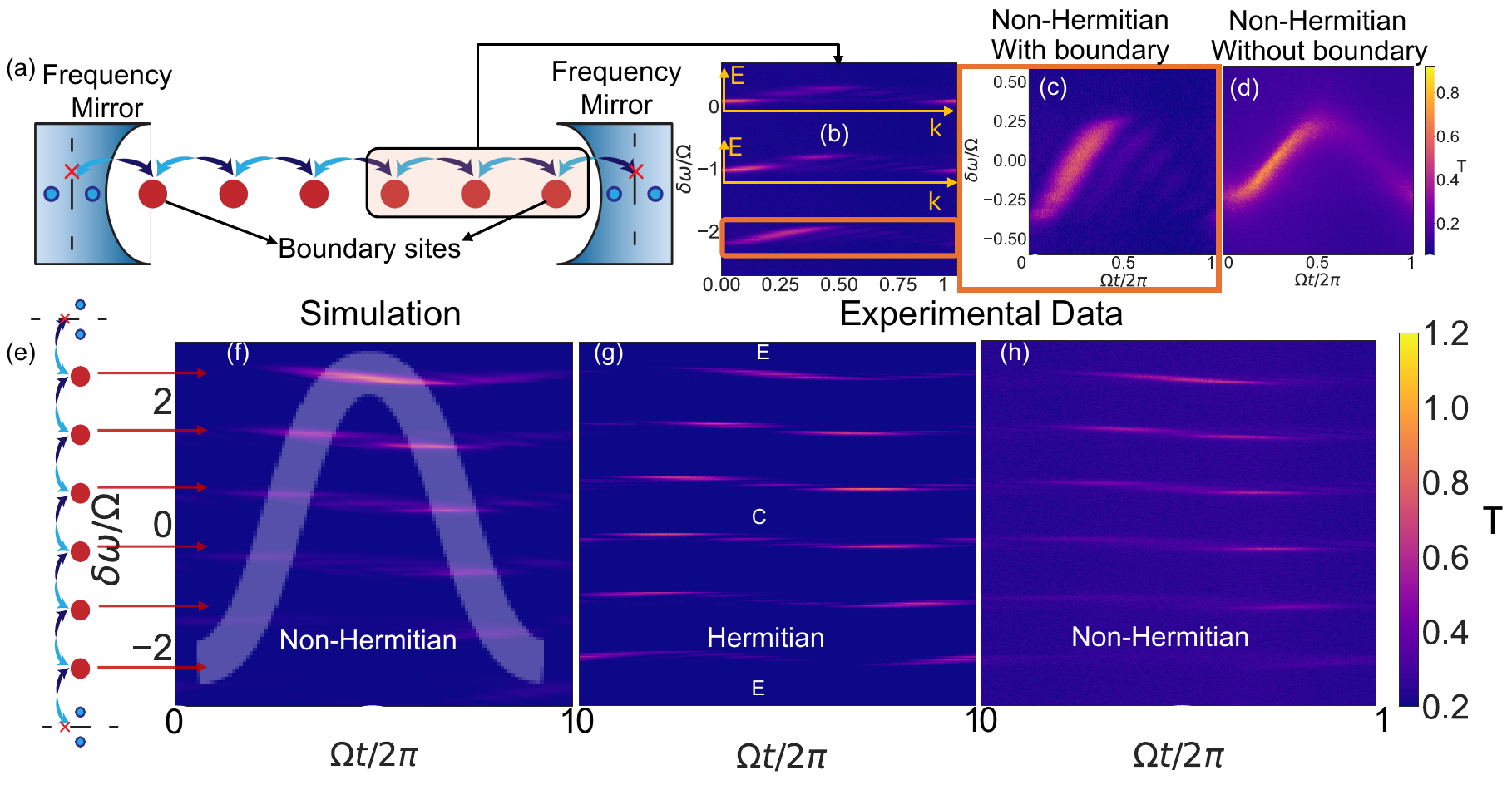}
    \caption{(a) The 1-D finite lattice in frequency has two frequency mirrors in a ring resonator with modest finesse, leading to a ``weakly confining frequency cavity" in which an excitation can propagate through lattice site couplings and reflect from the boundaries back into the lattice. 
    Excitation of the red highlighted region in (a) reveal the non-Hermitian lattice bands in (b). (c) and (d) are the non Hermitian bands obtained with and without frequency mirrors re-centered about 0. Arrows from (e) indicate the band structure obtained with the excitation of a particular site of the lattice. (f) Scattering matrix simulation of Non-Hermitian bands. An overall modulation of the brightness of the bands across the entire lattice due to the strong boundary is highlighted with the transparent sinusoid overlaid on the plot. (g) Experimentally measured Hermitian band structure of an entire 6-site lattice.  E denotes an excitation at the site closest to the edge of the lattice and C denotes an excitation of the central lattice sites. (h) Experimentally measured Non-Hermitian band structure of the 6 site lattice.}
    \label{fig:Bands}
\end{figure*}

\begin{figure*}[!htbp]
    \centering
    \includegraphics[width = \textwidth]{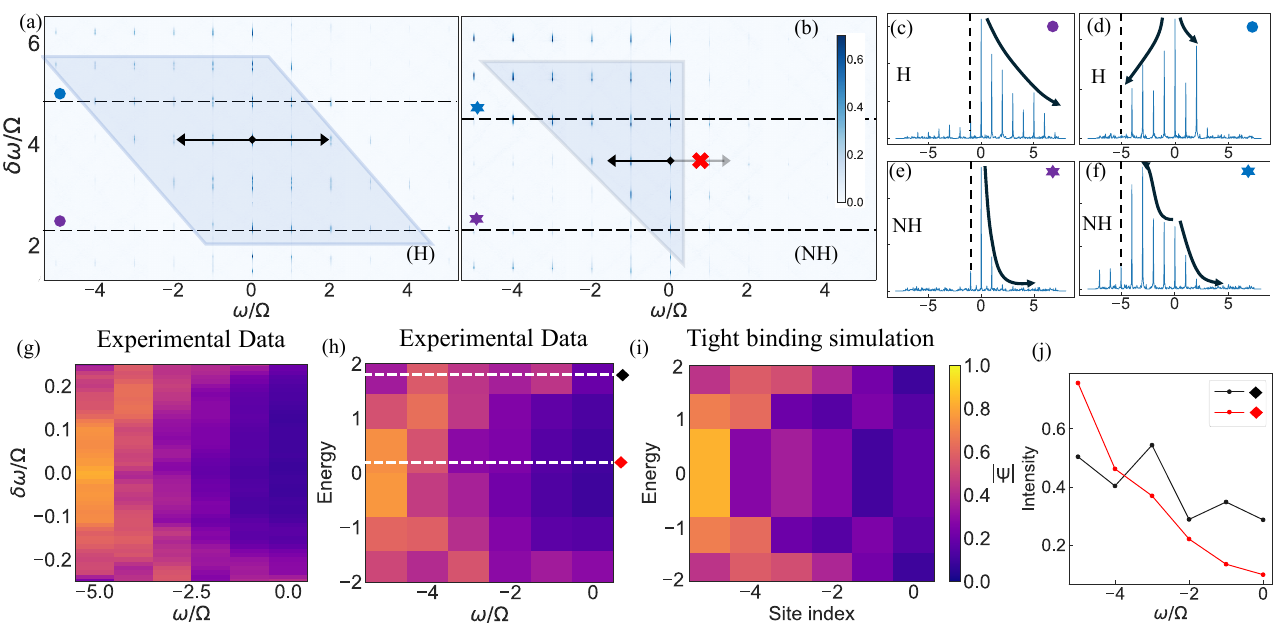}
    \caption{Steady state occupations of a slightly larger bounded 7-site (a) Hermitian (H) lattice and (b) non-Hermitian (NH) lattice, to better illustrate the effects of edge and central excitations. (c), (d):  line cuts for the Hermitian lattice. (e), (f): line cuts for the non-Hermitian lattice. (a) The highlighted region represents the finite Hermitian lattice across each detuning, where the occupation profile is symmetric. (b) The highlighted region shows the asymmetry pushing the excitation towards one direction across all detunings, resulting in a left-skewed occupation profile. Linecuts (c) and (e) represent the site next to the boundary in the Hermitian and non-Hermitian cases respectively. Linecuts (d) and (f) represent the site next to the central site of the lattice in the Hermitian and non-Hermitian cases respectively. Dotted black lines in line cuts (c)-(f) indicate the location of the boundary. The curved lines highlight the general envelope of the distribution from the site of excitation with the arrow in the direction of propagation. (g) Band resolved occupation profile of the original 6-site Non-Hermitian lattice plotted after normalizing the measured site amplitudes for uniform loss with an exponential envelope reveals a set of modes resembling the eigen mode amplitudes of the Hatano-Nelson lattice under OBC. (h) We select the eigen energies corresponding to theoretical OBC eigenvalues after normalization for a direct comparison of the experimentally extracted eigen modes and (i) theoretical eigen mode distribution computed using tight binding simulations. (j) Linecuts from (h) depicting trends described in Fig \ref{fig:Schematic}(e).
    }
    \label{fig:eigen}
\end{figure*}

As the Hamiltonian is non-Hermitian, the energy eigenvalues are complex and the winding of these eigenvalues in the complex energy plane defines the lattice topology. Under periodic boundary conditions, the eigenvalues trace an ellipse in the complex plane, whereas under open boundary conditions (OBC), the spectrum becomes strictly real \cite{martinez_alvarez_non-hermitian_2018, yao_edge_2018}, causing the ellipse to collapse onto a line along the real axis (Fig \ref{fig:Schematic}(d)). 
A physical consequence of the non-Hermitian Hamiltonian under OBC is the non-Hermitian skin effect. This manifests in the PBC bulk modes becoming exponentially localized at one of the boundaries of the OBC lattice. This is induced by the non-reciprocal couplings, thereby uniquely selecting which edge accumulates the eigenmodes. An example of such localized eigenmodes is depicted in the blue curves of Fig.~\ref{fig:Schematic}(e). The curves represent the mode amplitudes of eigenenergies near zero (light blue dots in Fig.~\ref{fig:Schematic}(d))  and at the maximum eigenenergy (dark blue dots in Fig.~\ref{fig:Schematic}(d)) of the Hatano-Nelson lattice under OBC. In both cases, the mode amplitude decays exponentially away from the selected edge. Additionally, eigenmodes closer to the extremal eigenenergies exhibit a slight inward shift in their peak amplitude away from the edge. We demonstrate both these physics effects in experiments (Fig.~\ref{fig:eigen}h) and their implications on the band structure~(Fig.~\ref{fig:Bands}b).

We experimentally realize the Hatano-Nelson Hamiltonian in the synthetic frequency dimension of a fiber ring resonator with a free spectral range (FSR) of \(13.595\) MHz as shown in Fig.~\ref{fig:Schematic}. Non‑reciprocal couplings are implemented by phase and amplitude modulation at the cavity FSR, with their strengths and relative phase electronically controlled to set the effective wavefunction phase and Hamiltonian parameters. An auxiliary ring that is $7\times$ smaller than the main ring is strongly coupled with a $50:50$ splitter, introducing a boundary with a large mode splitting of $\sim$8 MHz at every 7$^{\rm th}$ resonance of the main ring (Fig.~\ref{fig:Schematic}b). This mode splitting distorts the local resonance spacing and thereby suppresses further FSR‑mediated coupling, effectively terminating the synthetic lattice \cite{dutt_creating_2022,hu_mirror-induced_2022}. The large $8$-MHz splitting is vital to ensure that the nonreciprocal amplification of the underlying lattice is contained by the imposed boundary conditions.  From here on we will refer to the split mode as the ``frequency mirror" (Fig.~\ref{fig:Bands}a) and sites adjacent to the split modes as the ``boundary sites". The size of the lattice can be easily reconfigured by adjusting the length of the auxiliary ring.

As stated earlier, a unique advantage of our platform is the ability to simultaneously measure both in $k$-space (Fig.~\ref{fig:Bands}) and in lattice space (Fig.~\ref{fig:eigen}), with and without OBC,  which is particularly important for characterizing non-Hermitian signatures. Corresponding measurements for the Hermitian counterpart of the Hamiltonian further highlight the distinct nature of these effects. Together, these measurements provide a comprehensive picture of the physical phenomena enabled by NH Hamiltonians, which we elucidate below.

We apply the band structure spectroscopy technique~\cite{dutt_experimental_2019-1} --- widely used for Hamiltonian characterization in unbounded frequency lattices \cite{dutt_single_2020,wang_generating_2021,wang_topological_2021,li_dynamic_2021,balcytis_synthetic_2022,senanian_programmable_2023,li_direct_2023, cheng_multidimensional_2023, dinh_reconfigurable_2024, pellerin_wave-function_2024,chenier_quantized_2024, Wang_versatile_2025, ye_construction_2025} --- to our finite NH lattice.
Briefly reviewing the band structure spectroscopy technique: the Hamiltonian is formulated in the frequency basis of the resonator, making time the corresponding quasi-momentum $(k)$, and the laser detuning maps to the quasi-energy $(E)$. This mapping enables us to obtain the band structure directly from the time-resolved transmission of the resonator system measured with a high-speed oscilloscope (\(20\) GS/s). In our work, we engineer small 6-site and 7-site lattices to make boundary effects more pronounced. A salient feature of the bands in the finite frequency synthetic lattice is the presence of discrete brightness variations. One can draw a direct correspondence between the quasi-energy of each band and the eigen-energy, which is periodic as the lattice is engineered via Floquet modulation. 
Each band measures $E$ vs $k$ for the corresponding single site excitation, and the brightness of the signal indicates the projection of the excited site occupation onto the corresponding eigenmode at that detuning/eigenenergy. This is more generally applicable to frequency synthetic dimensions, and studies have used similar variations in brightness to perform wavefunction tomography on unbounded dimer lattices \cite{pellerin_wave-function_2024}. 

The discretized band structure of a finite 1-D Hermitian lattice has been demonstrated previously \cite{dutt_creating_2022}, and we verify this behavior in our system Fig \ref{fig:Bands}(e). We measure asymmetric non-Hermitian bands and contrast the fringing behavior in the finite lattice bands of Fig.~\ref{fig:Bands}(c) with the continuous bands of the unbounded lattice in Fig.~\ref{fig:Bands}(d), owing to the frequency mirrors.
For the NH case, we see that when we excite at the edge (Fig.~\ref{fig:Bands}(b) bottom band) the coefficients are larger for all the modes, exhibiting almost uniform brightness in the lobes, whereas exciting toward the center/bulk (Fig.~\ref{fig:Bands}(b) top), the lobes in the middle go dimmer while the top and bottom lobes (larger positive or negative eigenenergies) remain bright. 
This connects to the tight-binding lattice picture of the larger eigenvalue modes having more contribution from the bulk (Fig.~\ref{fig:Schematic}(e)), and further shows why the bands grow dimmer overall in the non-Hermitian case when moving away  from the direction of localization (Fig.~\ref{fig:Bands}(c, d, f), top to bottom). 
Moreover, due to the use of a large fiber-ring resonator with modest finesse, we operate in the regime of a ``weakly confining frequency cavity", quantified by the loss per hop and the hopping strength \cite{hu_mirror-induced_2022}. We also see the effect of reflection from either boundary as an overall sinusoidal variation in the bands when excited at each site, visible in simulated and experimentally obtained bands in Fig.~\ref{fig:Bands}(f) - (h).

The trends observed in the bands, although quite intuitive when understood from the perspective of frequency synthetic lattices, only offer an indirect probe into the eigenmodes of the system. We are thus motivated to confirm these claims with direct site-resolved probes of the corresponding eigenmodes, as $k$ in this finite lattice does not represent a good `quantum' number. For low-noise direct site-occupation measurements, we employ our previously proposed cascaded optical-electrical heterodyne scheme \cite{sridhar_measuring_2025} using an RF spectrum analyzer with a \(165\) MHz instantaneous IQ bandwidth. This enables one-shot measurements capturing nearly 2 full finite lattices ($\sim$ 12 sites). By sweeping the input laser across the entire lattice, we obtain site occupation measurements that reveal steady state populations across the simulated lattice.

\begin{figure*}[ht!]
    \centering
    \includegraphics[width = \textwidth]{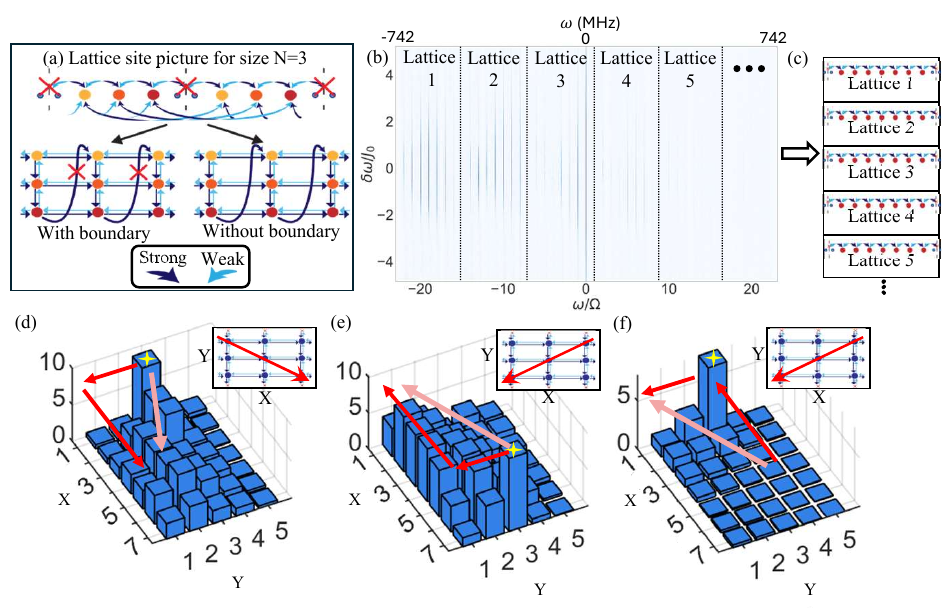}
    \caption{(a) 2-D lattice construction from finite lattice chain. The 2-D lattice on the left does not have the twisted coupling while the 2-D lattice on the right does (constructed with the same couplings but without a boundary).  (b) Experimentally measured frequency occupations. The dotted lines indicate the location of the boundary splitting that defines each finite lattice of the chain. (c) Stacking these lattices vertically aids visualization of the two dimensional structure as depicted in plots (d)-(f). (d)-(f) Experimental measurements of steady state occupations in the constructed 2-D lattices, measured at $\delta\omega/\Omega=0$. Yellow star indicates the site of excitation. Red arrows indicate the directions of NH asymmetry in each dimension. Light red arrows highlight the overall direction of 2-D asymmetry. The X dimension represents the finite lattice between boundaries (Each box in (c)) and the Y dimension is unbounded, describing the connectivity between lattices. Insets provide a 2-D view of the overall direction of asymmetry in the 2-D lattice. (d) Site occupations with stronger coupling in the right (+X) and bottom (-Y) directions, showing the expected accumulation towards high X and low Y-valued sites. (e) Same as (d) but with stronger coupling towards left (-X). The accumulation now switches towards lower X-valued sites. (f) Site occupations with the same coupling parameters as (e) with excitation of the other end of the lattice. The asymmetry pushes the photons towards the boundary and hence localizes it in one dimension dimensionally reducing the 2-D lattice to an occupation of an unbounded 1-D lattice along the Y direction.
    }
    \label{fig:nh2d}
\end{figure*}

We observe strong boundary localization characteristic of skin modes in a 1-D NH frequency synthetic lattice. 
Fig. \ref{fig:eigen}(a,b) compares bounded 7-site 1-D Hermitian and non-Hermitian lattices using the aforementioned cascaded heterodyne technique. The highlighted region in Fig.~\ref{fig:eigen}(a) indicates the symmetric coupling of the Hermitian lattice as we sweep the detuning, while in the non-Hermitian case (Fig.~\ref{fig:eigen}(b)), there is a strong asymmetry towards the left that pushes the excitation away from the center. Here, propagation towards the right is forbidden. An excitation adjacent to the boundary in the non-Hermitian (Fig.~\ref{fig:eigen}(e)) case exhibits sharper localization, as evidenced by stronger decay from asymmetric couplings than the slower, uniform-loss-driven decay seen in the Hermitian (Fig. \ref{fig:eigen}(c)) lattice. Excitation of a site near the center of the Hermitian (Fig. \ref{fig:eigen}(d)) lattice symmetrically decays on both sides, while the non-Hermitian lattice (Fig. \ref{fig:eigen}(f)) exhibits markedly different behavior: amplification in the direction of asymmetry. These observations are due to the presence of an NH skin mode at the boundary and directionality achieved across the entire lattice.

We further demonstrate that the low-noise frequency measurement of the lattice directly reveals the eigenmodes -- the occupation profiles vary significantly with detuning (or eigenenergy), a characteristic signature of the underlying lattice. Slowly sweeping the input across each site, we obtain eigenenergy/band-resolved site occupation measurements, which we term ``eigenmode spectroscopy". To analyze this, we compare the measured occupation profiles of a 6-site lattice with tight-binding simulations of the Hatano-Nelson lattice under OBC (Fig.~\ref{fig:Schematic}(e)). Based on the bands measured in Fig. 2, the singular bright lobe at the boundary excitation indicates uniform overlap with all the eigenmodes. We therefore perform a fine sweep across this site, marked with the purple star in Fig.~\ref{fig:eigen}(b), with a zoom-in on one of  the NHSE modes in Fig.~\ref{fig:eigen}(e). The experimental data in Fig. \ref{fig:eigen}(g) shows the stacked occupations from each detuning, where we can now normalize the detuning axis to the eigenenergy range and pick out the corresponding eigenmode profiles. We also account for the uniform loss in the lattice by normalizing the measured site amplitudes with an exponential envelope, specified by the cavity round-trip loss and hopping strength \cite{zhang_broadband_2019,hu_mirror-induced_2022}. This produces the eigenmode spectrum in Fig.~ \ref{fig:eigen}(h), which we contrast with the tight-binding calculation in Fig.~ \ref{fig:eigen}(i). We observe the trends highlighted earlier in the eigenmode profiles (Fig.~\ref{fig:Schematic}(e)) corresponding to energy eigenvalues close to zero and near the spectral extrema. The line cuts shown in Figure~\ref{fig:eigen}(j) confirm both the localization of the eigenmodes near one of the boundaries and a slight inward shift of the peak position for the eigenmode associated with an extremal energy eigenvalue. 

\textit{Extending to higher-dimensional NH lattices:} So far, we investigate the effects caused by finiteness in a single 1-D lattice. By adding a long-range coupling between modes of different finite lattices we can construct 2-D lattices \cite{senanian_programmable_2023,cheng_multidimensional_2023}. High-dimensional non-Hermitian lattices offer a rich landscape of phenomena, including geometry-dependent boundary sensitivity \cite{Yang_tailoring_2025}, higher-order exceptional points \cite{Montag_symmetry-induced_2024}, and engineered localization such as surface modes and corner modes \cite{zhao_non-hermitian_2019,Zou_observation_2021,zheng_dynamic_2024}. These phenomena arise from an intricate interplay between dimensionality, non-Hermiticity, and topology. 

Motivated by the interest in higher-dimensional topology, the incorporation of long-range couplings in a 1-D synthetic lattice has been shown to mimic certain physics effects expected in higher-D lattices, such as two-dimensional transport \cite{senanian_programmable_2023} or Haldane-like models \cite{yuan_synthetic-space_2018, chenier_quantized_2024}. However, such constructions can be reduced to an equivalent 1-D representation. In other words, the reciprocal space can still be spanned by a 1-D Brillouin zone indexed by a single quasimomentum $k$ due to the underlying 1-D translational symmetry that remains manifest. The specific effect introduced is the presence of twisted boundaries, whereby the lattice can be visualized on a tube ~\cite{wang_multidimensional_2020, yuan_synthetic-space_2018}. 

Our strong boundary explicitly breaks 1-D translational symmetry, enabling the construction of clean, \textit{genuine} 2-D rectangular lattices that are irreducible to a 1-D lattice Hamiltonian (Fig \ref{fig:nh2d}(a)). 
As described before, the auxiliary ring creates a boundary via a mode splitting at every $N$ modes of the main ring. This creates a series of finite lattices with the boundaries of each separated by the frequency mirror. 

To verify 2-D lattice construction, we employ the site-resolved measurements as in the 1-D case. By virtue of our experiment's passive stability, we can sweep the radio-frequency reference and obtain site occupations by performing optical heterodyne at the drop port of our cavity. We hence sequentially acquire site-occupation data for up to 50 sites by stitching a plurality of frequency segments together. After identifying the position of the boundary in the data, the intensities from the central eigenenergy $(\delta\omega/\Omega=0)$ can be stacked in a two dimensional format. This process is highlighted in Fig.~\ref{fig:nh2d}(b,c). 

As a final demonstration, we assemble all of the capabilities developed in this paper to measure directionally tunable transport across 2-D frequency space for the first time. 
By maintaining the asymmetry along the Y-direction toward negative Y, we demonstrate tunable directional transport in the lattice along the southeast (Fig.~\ref{fig:nh2d}(d)) and southwest (Fig.~\ref{fig:nh2d}(e)) directions. A distinct configuration is presented in Fig.~\ref{fig:nh2d}(f), which employs the same asymmetry parameters as Fig.~\ref{fig:nh2d}(e) but with the excitation at a different site (the opposite X-boundary site). This results in the localization of the lattice population along the X-boundary while unbounded along Y, corresponding to edge localization in the two-dimensional frequency space \cite{Zhao_two-dimensional_2025}. 

These results underscore the remarkable versatility of the platform, establishing it as a powerful and highly controllable testbed for the realization and exploration of engineered non-Hermitian Hamiltonians and higher-dimensional non-Hermitian phenomena.

\textit{Conclusions}: In this work we have provided one of the first experimental demonstrations of the non-Hermitian skin effect in synthetic frequency dimensions by showcasing exponential localization at the boundary and eigenmode spectroscopy under open boundary conditions. Furthermore we added asymmetric longer-range couplings to demonstrate for the first time genuine 2-D lattices in frequency space, highlighting tunable directionality of transport. The versatility of the platform in conjunction with the tools we developed for Hamiltonian construction and detection make this a leading platform for Hamiltonian simulation and sensing applications. Furthermore, integrating this system onto a chip with high-finesse cavities can lead to more exotic ``frequency cavity" effects, with potential quantum sensing implications \cite{hu_mirror-induced_2022}.

\section*{Acknowledgements}
This work was supported by Northrop Grumman and Leidos, and by an NSF CAREER award (2340835). 

\bibliography{avik_zotero_library, mainbib}

\providecommand{\noopsort}[1]{}\providecommand{\singleletter}[1]{#1}%
\begin{thebibliography}{49}%
\makeatletter
\providecommand \@ifxundefined [1]{%
 \@ifx{#1\undefined}
}%
\providecommand \@ifnum [1]{%
 \ifnum #1\expandafter \@firstoftwo
 \else \expandafter \@secondoftwo
 \fi
}%
\providecommand \@ifx [1]{%
 \ifx #1\expandafter \@firstoftwo
 \else \expandafter \@secondoftwo
 \fi
}%
\providecommand \natexlab [1]{#1}%
\providecommand \enquote  [1]{``#1''}%
\providecommand \bibnamefont  [1]{#1}%
\providecommand \bibfnamefont [1]{#1}%
\providecommand \citenamefont [1]{#1}%
\providecommand \href@noop [0]{\@secondoftwo}%
\providecommand \href [0]{\begingroup \@sanitize@url \@href}%
\providecommand \@href[1]{\@@startlink{#1}\@@href}%
\providecommand \@@href[1]{\endgroup#1\@@endlink}%
\providecommand \@sanitize@url [0]{\catcode `\\12\catcode `\$12\catcode `\&12\catcode `\#12\catcode `\^12\catcode `\_12\catcode `\%12\relax}%
\providecommand \@@startlink[1]{}%
\providecommand \@@endlink[0]{}%
\providecommand \url  [0]{\begingroup\@sanitize@url \@url }%
\providecommand \@url [1]{\endgroup\@href {#1}{\urlprefix }}%
\providecommand \urlprefix  [0]{URL }%
\providecommand \Eprint [0]{\href }%
\providecommand \doibase [0]{https://doi.org/}%
\providecommand \selectlanguage [0]{\@gobble}%
\providecommand \bibinfo  [0]{\@secondoftwo}%
\providecommand \bibfield  [0]{\@secondoftwo}%
\providecommand \translation [1]{[#1]}%
\providecommand \BibitemOpen [0]{}%
\providecommand \bibitemStop [0]{}%
\providecommand \bibitemNoStop [0]{.\EOS\space}%
\providecommand \EOS [0]{\spacefactor3000\relax}%
\providecommand \BibitemShut  [1]{\csname bibitem#1\endcsname}%
\let\auto@bib@innerbib\@empty
\bibitem [{\citenamefont {Hatano}\ and\ \citenamefont {Nelson}(1996)}]{hatano_localization_1996}%
  \BibitemOpen
  \bibfield  {author} {\bibinfo {author} {\bibfnamefont {N.}~\bibnamefont {Hatano}}\ and\ \bibinfo {author} {\bibfnamefont {D.~R.}\ \bibnamefont {Nelson}},\ }\bibfield  {title} {\bibinfo {title} {Localization {Transitions} in {Non}-{Hermitian} {Quantum} {Mechanics}},\ }\href {https://doi.org/10.1103/PhysRevLett.77.570} {\bibfield  {journal} {\bibinfo  {journal} {Phys. Rev. Lett.}\ }\textbf {\bibinfo {volume} {77}},\ \bibinfo {pages} {570} (\bibinfo {year} {1996})}\BibitemShut {NoStop}%
\bibitem [{\citenamefont {Martinez~Alvarez}\ \emph {et~al.}(2018)\citenamefont {Martinez~Alvarez}, \citenamefont {Barrios~Vargas},\ and\ \citenamefont {Foa~Torres}}]{martinez_alvarez_non-hermitian_2018}%
  \BibitemOpen
  \bibfield  {author} {\bibinfo {author} {\bibfnamefont {V.~M.}\ \bibnamefont {Martinez~Alvarez}}, \bibinfo {author} {\bibfnamefont {J.~E.}\ \bibnamefont {Barrios~Vargas}},\ and\ \bibinfo {author} {\bibfnamefont {L.~E.~F.}\ \bibnamefont {Foa~Torres}},\ }\bibfield  {title} {\bibinfo {title} {Non-hermitian robust edge states in one dimension: Anomalous localization and eigenspace condensation at exceptional points},\ }\href {https://doi.org/10.1103/PhysRevB.97.121401} {\bibfield  {journal} {\bibinfo  {journal} {Phys. Rev. B}\ }\textbf {\bibinfo {volume} {97}},\ \bibinfo {pages} {121401(R)} (\bibinfo {year} {2018})}\BibitemShut {NoStop}%
\bibitem [{\citenamefont {Yao}\ and\ \citenamefont {Wang}(2018)}]{yao_edge_2018}%
  \BibitemOpen
  \bibfield  {author} {\bibinfo {author} {\bibfnamefont {S.}~\bibnamefont {Yao}}\ and\ \bibinfo {author} {\bibfnamefont {Z.}~\bibnamefont {Wang}},\ }\bibfield  {title} {\bibinfo {title} {Edge states and topological invariants of non-hermitian systems},\ }\href {https://doi.org/10.1103/PhysRevLett.121.086803} {\bibfield  {journal} {\bibinfo  {journal} {Phys. Rev. Lett.}\ }\textbf {\bibinfo {volume} {121}},\ \bibinfo {pages} {086803} (\bibinfo {year} {2018})}\BibitemShut {NoStop}%
\bibitem [{\citenamefont {Okuma}\ \emph {et~al.}(2020)\citenamefont {Okuma}, \citenamefont {Kawabata}, \citenamefont {Shiozaki},\ and\ \citenamefont {Sato}}]{okuma_topological_2020}%
  \BibitemOpen
  \bibfield  {author} {\bibinfo {author} {\bibfnamefont {N.}~\bibnamefont {Okuma}}, \bibinfo {author} {\bibfnamefont {K.}~\bibnamefont {Kawabata}}, \bibinfo {author} {\bibfnamefont {K.}~\bibnamefont {Shiozaki}},\ and\ \bibinfo {author} {\bibfnamefont {M.}~\bibnamefont {Sato}},\ }\bibfield  {title} {\bibinfo {title} {Topological {Origin} of {Non}-{Hermitian} {Skin} {Effects}},\ }\href {https://doi.org/10.1103/PhysRevLett.124.086801} {\bibfield  {journal} {\bibinfo  {journal} {Phys. Rev. Lett.}\ }\textbf {\bibinfo {volume} {124}},\ \bibinfo {pages} {086801} (\bibinfo {year} {2020})}\BibitemShut {NoStop}%
\bibitem [{\citenamefont {McDonald}\ and\ \citenamefont {Clerk}(2020)}]{mcdonald_exponentially-enhanced_2020}%
  \BibitemOpen
  \bibfield  {author} {\bibinfo {author} {\bibfnamefont {A.}~\bibnamefont {McDonald}}\ and\ \bibinfo {author} {\bibfnamefont {A.~A.}\ \bibnamefont {Clerk}},\ }\bibfield  {title} {\bibinfo {title} {Exponentially-enhanced quantum sensing with non-{Hermitian} lattice dynamics},\ }\href {https://doi.org/10.1038/s41467-020-19090-4} {\bibfield  {journal} {\bibinfo  {journal} {Nat Commun}\ }\textbf {\bibinfo {volume} {11}},\ \bibinfo {pages} {5382} (\bibinfo {year} {2020})}\BibitemShut {NoStop}%
\bibitem [{\citenamefont {Liu}\ \emph {et~al.}(2021)\citenamefont {Liu}, \citenamefont {Shao}, \citenamefont {Ma}, \citenamefont {Zhang}, \citenamefont {You}, \citenamefont {Wu}, \citenamefont {Xiang}, \citenamefont {Cui},\ and\ \citenamefont {Zhang}}]{Liu_non-hermitian_2021}%
  \BibitemOpen
  \bibfield  {author} {\bibinfo {author} {\bibfnamefont {S.}~\bibnamefont {Liu}}, \bibinfo {author} {\bibfnamefont {R.}~\bibnamefont {Shao}}, \bibinfo {author} {\bibfnamefont {S.}~\bibnamefont {Ma}}, \bibinfo {author} {\bibfnamefont {L.}~\bibnamefont {Zhang}}, \bibinfo {author} {\bibfnamefont {O.}~\bibnamefont {You}}, \bibinfo {author} {\bibfnamefont {H.}~\bibnamefont {Wu}}, \bibinfo {author} {\bibfnamefont {Y.~J.}\ \bibnamefont {Xiang}}, \bibinfo {author} {\bibfnamefont {T.~J.}\ \bibnamefont {Cui}},\ and\ \bibinfo {author} {\bibfnamefont {S.}~\bibnamefont {Zhang}},\ }\bibfield  {title} {\bibinfo {title} {Non-hermitian skin effect in a non-hermitian electrical circuit},\ }\bibfield  {journal} {\bibinfo  {journal} {Research}\ }\textbf {\bibinfo {volume} {2021}},\ \href {https://doi.org/10.34133/2021/5608038} {10.34133/2021/5608038} (\bibinfo {year} {2021})\BibitemShut {NoStop}%
\bibitem [{\citenamefont {Yuan}\ \emph {et~al.}(2023)\citenamefont {Yuan}, \citenamefont {Zhang}, \citenamefont {Zhou}, \citenamefont {Wang}, \citenamefont {Pan}, \citenamefont {Feng}, \citenamefont {Sun},\ and\ \citenamefont {Zhang}}]{Yuan_non-hermitian_2023}%
  \BibitemOpen
  \bibfield  {author} {\bibinfo {author} {\bibfnamefont {H.}~\bibnamefont {Yuan}}, \bibinfo {author} {\bibfnamefont {W.}~\bibnamefont {Zhang}}, \bibinfo {author} {\bibfnamefont {Z.}~\bibnamefont {Zhou}}, \bibinfo {author} {\bibfnamefont {W.}~\bibnamefont {Wang}}, \bibinfo {author} {\bibfnamefont {N.}~\bibnamefont {Pan}}, \bibinfo {author} {\bibfnamefont {Y.}~\bibnamefont {Feng}}, \bibinfo {author} {\bibfnamefont {H.}~\bibnamefont {Sun}},\ and\ \bibinfo {author} {\bibfnamefont {X.}~\bibnamefont {Zhang}},\ }\bibfield  {title} {\bibinfo {title} {Non‐hermitian topolectrical circuit sensor with high sensitivity},\ }\bibfield  {journal} {\bibinfo  {journal} {Advanced Science}\ }\textbf {\bibinfo {volume} {10}},\ \href {https://doi.org/10.1002/advs.202301128} {10.1002/advs.202301128} (\bibinfo {year} {2023})\BibitemShut {NoStop}%
\bibitem [{\citenamefont {Parto}\ \emph {et~al.}(2025)\citenamefont {Parto}, \citenamefont {Leefmans}, \citenamefont {Williams}, \citenamefont {Gray},\ and\ \citenamefont {Marandi}}]{Parto_enhanced_2025}%
  \BibitemOpen
  \bibfield  {author} {\bibinfo {author} {\bibfnamefont {M.}~\bibnamefont {Parto}}, \bibinfo {author} {\bibfnamefont {C.}~\bibnamefont {Leefmans}}, \bibinfo {author} {\bibfnamefont {J.}~\bibnamefont {Williams}}, \bibinfo {author} {\bibfnamefont {R.~M.}\ \bibnamefont {Gray}},\ and\ \bibinfo {author} {\bibfnamefont {A.}~\bibnamefont {Marandi}},\ }\bibfield  {title} {\bibinfo {title} {Enhanced sensitivity via non-hermitian topology},\ }\bibfield  {journal} {\bibinfo  {journal} {Light: Science \& Applications}\ }\textbf {\bibinfo {volume} {14}},\ \href {https://doi.org/10.1038/s41377-024-01667-z} {10.1038/s41377-024-01667-z} (\bibinfo {year} {2025})\BibitemShut {NoStop}%
\bibitem [{\citenamefont {Wang}\ \emph {et~al.}(2025{\natexlab{a}})\citenamefont {Wang}, \citenamefont {Fu}, \citenamefont {Ye}, \citenamefont {He}, \citenamefont {Deng}, \citenamefont {Lu}, \citenamefont {Ke},\ and\ \citenamefont {Liu}}]{wang_observation_2025}%
  \BibitemOpen
  \bibfield  {author} {\bibinfo {author} {\bibfnamefont {Q.}~\bibnamefont {Wang}}, \bibinfo {author} {\bibfnamefont {Z.}~\bibnamefont {Fu}}, \bibinfo {author} {\bibfnamefont {L.}~\bibnamefont {Ye}}, \bibinfo {author} {\bibfnamefont {H.}~\bibnamefont {He}}, \bibinfo {author} {\bibfnamefont {W.}~\bibnamefont {Deng}}, \bibinfo {author} {\bibfnamefont {J.}~\bibnamefont {Lu}}, \bibinfo {author} {\bibfnamefont {M.}~\bibnamefont {Ke}},\ and\ \bibinfo {author} {\bibfnamefont {Z.}~\bibnamefont {Liu}},\ }\bibfield  {title} {\bibinfo {title} {Observation of supersensitivity of non-hermitian skin effect},\ }\href {https://doi.org/10.1103/8zfx-c7mw} {\bibfield  {journal} {\bibinfo  {journal} {Phys. Rev. Res.}\ }\textbf {\bibinfo {volume} {7}},\ \bibinfo {pages} {L042057} (\bibinfo {year} {2025}{\natexlab{a}})}\BibitemShut {NoStop}%
\bibitem [{\citenamefont {Ghatak}\ \emph {et~al.}(2020)\citenamefont {Ghatak}, \citenamefont {Brandenbourger}, \citenamefont {van Wezel},\ and\ \citenamefont {Coulais}}]{ghatak_observation_2020}%
  \BibitemOpen
  \bibfield  {author} {\bibinfo {author} {\bibfnamefont {A.}~\bibnamefont {Ghatak}}, \bibinfo {author} {\bibfnamefont {M.}~\bibnamefont {Brandenbourger}}, \bibinfo {author} {\bibfnamefont {J.}~\bibnamefont {van Wezel}},\ and\ \bibinfo {author} {\bibfnamefont {C.}~\bibnamefont {Coulais}},\ }\bibfield  {title} {\bibinfo {title} {Observation of non-{Hermitian} topology and its bulk–edge correspondence in an active mechanical metamaterial},\ }\href {https://doi.org/10.1073/pnas.2010580117} {\bibfield  {journal} {\bibinfo  {journal} {Proceedings of the National Academy of Sciences}\ }\textbf {\bibinfo {volume} {117}},\ \bibinfo {pages} {29561} (\bibinfo {year} {2020})}\BibitemShut {NoStop}%
\bibitem [{\citenamefont {Zhang}\ \emph {et~al.}(2021)\citenamefont {Zhang}, \citenamefont {Tian}, \citenamefont {Jiang}, \citenamefont {Lu},\ and\ \citenamefont {Chen}}]{Zhang_observation_2021}%
  \BibitemOpen
  \bibfield  {author} {\bibinfo {author} {\bibfnamefont {X.}~\bibnamefont {Zhang}}, \bibinfo {author} {\bibfnamefont {Y.}~\bibnamefont {Tian}}, \bibinfo {author} {\bibfnamefont {J.-H.}\ \bibnamefont {Jiang}}, \bibinfo {author} {\bibfnamefont {M.-H.}\ \bibnamefont {Lu}},\ and\ \bibinfo {author} {\bibfnamefont {Y.-F.}\ \bibnamefont {Chen}},\ }\bibfield  {title} {\bibinfo {title} {Observation of higher-order non-hermitian skin effect},\ }\bibfield  {journal} {\bibinfo  {journal} {Nature Communications}\ }\textbf {\bibinfo {volume} {12}},\ \href {https://doi.org/10.1038/s41467-021-25716-y} {10.1038/s41467-021-25716-y} (\bibinfo {year} {2021})\BibitemShut {NoStop}%
\bibitem [{\citenamefont {Gu}\ \emph {et~al.}(2022)\citenamefont {Gu}, \citenamefont {Gao}, \citenamefont {Xue}, \citenamefont {Li}, \citenamefont {Su},\ and\ \citenamefont {Zhu}}]{Gu_transient_2022}%
  \BibitemOpen
  \bibfield  {author} {\bibinfo {author} {\bibfnamefont {Z.}~\bibnamefont {Gu}}, \bibinfo {author} {\bibfnamefont {H.}~\bibnamefont {Gao}}, \bibinfo {author} {\bibfnamefont {H.}~\bibnamefont {Xue}}, \bibinfo {author} {\bibfnamefont {J.}~\bibnamefont {Li}}, \bibinfo {author} {\bibfnamefont {Z.}~\bibnamefont {Su}},\ and\ \bibinfo {author} {\bibfnamefont {J.}~\bibnamefont {Zhu}},\ }\bibfield  {title} {\bibinfo {title} {Transient non-hermitian skin effect},\ }\bibfield  {journal} {\bibinfo  {journal} {Nature Communications}\ }\textbf {\bibinfo {volume} {13}},\ \href {https://doi.org/10.1038/s41467-022-35448-2} {10.1038/s41467-022-35448-2} (\bibinfo {year} {2022})\BibitemShut {NoStop}%
\bibitem [{\citenamefont {Helbig}\ \emph {et~al.}(2020)\citenamefont {Helbig}, \citenamefont {Hofmann}, \citenamefont {Imhof}, \citenamefont {Abdelghany}, \citenamefont {Kiessling}, \citenamefont {Molenkamp}, \citenamefont {Lee}, \citenamefont {Szameit}, \citenamefont {Greiter},\ and\ \citenamefont {Thomale}}]{Helbig_generalized_2020}%
  \BibitemOpen
  \bibfield  {author} {\bibinfo {author} {\bibfnamefont {T.}~\bibnamefont {Helbig}}, \bibinfo {author} {\bibfnamefont {T.}~\bibnamefont {Hofmann}}, \bibinfo {author} {\bibfnamefont {S.}~\bibnamefont {Imhof}}, \bibinfo {author} {\bibfnamefont {M.}~\bibnamefont {Abdelghany}}, \bibinfo {author} {\bibfnamefont {T.}~\bibnamefont {Kiessling}}, \bibinfo {author} {\bibfnamefont {L.~W.}\ \bibnamefont {Molenkamp}}, \bibinfo {author} {\bibfnamefont {C.~H.}\ \bibnamefont {Lee}}, \bibinfo {author} {\bibfnamefont {A.}~\bibnamefont {Szameit}}, \bibinfo {author} {\bibfnamefont {M.}~\bibnamefont {Greiter}},\ and\ \bibinfo {author} {\bibfnamefont {R.}~\bibnamefont {Thomale}},\ }\bibfield  {title} {\bibinfo {title} {Generalized bulk–boundary correspondence in non-hermitian topolectrical circuits},\ }\href {https://doi.org/10.1038/s41567-020-0922-9} {\bibfield  {journal} {\bibinfo  {journal} {Nature Physics}\ }\textbf {\bibinfo {volume} {16}},\ \bibinfo {pages} {747–750} (\bibinfo {year} {2020})}\BibitemShut {NoStop}%
\bibitem [{\citenamefont {Zhu}\ \emph {et~al.}(2023)\citenamefont {Zhu}, \citenamefont {Sun}, \citenamefont {Hughes},\ and\ \citenamefont {Bahl}}]{zhu_higher_2023}%
  \BibitemOpen
  \bibfield  {author} {\bibinfo {author} {\bibfnamefont {P.}~\bibnamefont {Zhu}}, \bibinfo {author} {\bibfnamefont {X.-Q.}\ \bibnamefont {Sun}}, \bibinfo {author} {\bibfnamefont {T.~L.}\ \bibnamefont {Hughes}},\ and\ \bibinfo {author} {\bibfnamefont {G.}~\bibnamefont {Bahl}},\ }\bibfield  {title} {\bibinfo {title} {Higher rank chirality and non-{Hermitian} skin effect in a topolectrical circuit},\ }\href {https://doi.org/10.1038/s41467-023-36130-x} {\bibfield  {journal} {\bibinfo  {journal} {Nat Commun}\ }\textbf {\bibinfo {volume} {14}},\ \bibinfo {pages} {720} (\bibinfo {year} {2023})}\BibitemShut {NoStop}%
\bibitem [{\citenamefont {Liang}\ \emph {et~al.}(2022)\citenamefont {Liang}, \citenamefont {Xie}, \citenamefont {Dong}, \citenamefont {Li}, \citenamefont {Li}, \citenamefont {Gadway}, \citenamefont {Yi},\ and\ \citenamefont {Yan}}]{Liang_dynamic_2022}%
  \BibitemOpen
  \bibfield  {author} {\bibinfo {author} {\bibfnamefont {Q.}~\bibnamefont {Liang}}, \bibinfo {author} {\bibfnamefont {D.}~\bibnamefont {Xie}}, \bibinfo {author} {\bibfnamefont {Z.}~\bibnamefont {Dong}}, \bibinfo {author} {\bibfnamefont {H.}~\bibnamefont {Li}}, \bibinfo {author} {\bibfnamefont {H.}~\bibnamefont {Li}}, \bibinfo {author} {\bibfnamefont {B.}~\bibnamefont {Gadway}}, \bibinfo {author} {\bibfnamefont {W.}~\bibnamefont {Yi}},\ and\ \bibinfo {author} {\bibfnamefont {B.}~\bibnamefont {Yan}},\ }\bibfield  {title} {\bibinfo {title} {Dynamic signatures of non-hermitian skin effect and topology in ultracold atoms},\ }\href {https://doi.org/10.1103/PhysRevLett.129.070401} {\bibfield  {journal} {\bibinfo  {journal} {Phys. Rev. Lett.}\ }\textbf {\bibinfo {volume} {129}},\ \bibinfo {pages} {070401} (\bibinfo {year} {2022})}\BibitemShut {NoStop}%
\bibitem [{\citenamefont {Zhao}\ \emph {et~al.}(2025)\citenamefont {Zhao}, \citenamefont {Wang}, \citenamefont {He}, \citenamefont {Poon}, \citenamefont {Pak}, \citenamefont {Liu}, \citenamefont {Ren}, \citenamefont {Liu},\ and\ \citenamefont {Jo}}]{Zhao_two-dimensional_2025}%
  \BibitemOpen
  \bibfield  {author} {\bibinfo {author} {\bibfnamefont {E.}~\bibnamefont {Zhao}}, \bibinfo {author} {\bibfnamefont {Z.}~\bibnamefont {Wang}}, \bibinfo {author} {\bibfnamefont {C.}~\bibnamefont {He}}, \bibinfo {author} {\bibfnamefont {T.~F.~J.}\ \bibnamefont {Poon}}, \bibinfo {author} {\bibfnamefont {K.~K.}\ \bibnamefont {Pak}}, \bibinfo {author} {\bibfnamefont {Y.-J.}\ \bibnamefont {Liu}}, \bibinfo {author} {\bibfnamefont {P.}~\bibnamefont {Ren}}, \bibinfo {author} {\bibfnamefont {X.-J.}\ \bibnamefont {Liu}},\ and\ \bibinfo {author} {\bibfnamefont {G.-B.}\ \bibnamefont {Jo}},\ }\bibfield  {title} {\bibinfo {title} {Two-dimensional non-hermitian skin effect in an ultracold fermi gas},\ }\href {https://doi.org/10.1038/s41586-024-08347-3} {\bibfield  {journal} {\bibinfo  {journal} {Nature}\ }\textbf {\bibinfo {volume} {637}},\ \bibinfo {pages} {565–573} (\bibinfo {year} {2025})}\BibitemShut {NoStop}%
\bibitem [{\citenamefont {Weidemann}\ \emph {et~al.}(2020)\citenamefont {Weidemann}, \citenamefont {Kremer}, \citenamefont {Helbig}, \citenamefont {Hofmann}, \citenamefont {Stegmaier}, \citenamefont {Greiter}, \citenamefont {Thomale},\ and\ \citenamefont {Szameit}}]{weidemann_topological_2020}%
  \BibitemOpen
  \bibfield  {author} {\bibinfo {author} {\bibfnamefont {S.}~\bibnamefont {Weidemann}}, \bibinfo {author} {\bibfnamefont {M.}~\bibnamefont {Kremer}}, \bibinfo {author} {\bibfnamefont {T.}~\bibnamefont {Helbig}}, \bibinfo {author} {\bibfnamefont {T.}~\bibnamefont {Hofmann}}, \bibinfo {author} {\bibfnamefont {A.}~\bibnamefont {Stegmaier}}, \bibinfo {author} {\bibfnamefont {M.}~\bibnamefont {Greiter}}, \bibinfo {author} {\bibfnamefont {R.}~\bibnamefont {Thomale}},\ and\ \bibinfo {author} {\bibfnamefont {A.}~\bibnamefont {Szameit}},\ }\bibfield  {title} {\bibinfo {title} {Topological funneling of light},\ }\href {https://doi.org/10.1126/science.aaz8727} {\bibfield  {journal} {\bibinfo  {journal} {Science}\ }\textbf {\bibinfo {volume} {368}},\ \bibinfo {pages} {311} (\bibinfo {year} {2020})}\BibitemShut {NoStop}%
\bibitem [{\citenamefont {Xiao}\ \emph {et~al.}(2020)\citenamefont {Xiao}, \citenamefont {Deng}, \citenamefont {Wang}, \citenamefont {Zhu}, \citenamefont {Wang}, \citenamefont {Yi},\ and\ \citenamefont {Xue}}]{Xiao_non-hermitian_2020}%
  \BibitemOpen
  \bibfield  {author} {\bibinfo {author} {\bibfnamefont {L.}~\bibnamefont {Xiao}}, \bibinfo {author} {\bibfnamefont {T.}~\bibnamefont {Deng}}, \bibinfo {author} {\bibfnamefont {K.}~\bibnamefont {Wang}}, \bibinfo {author} {\bibfnamefont {G.}~\bibnamefont {Zhu}}, \bibinfo {author} {\bibfnamefont {Z.}~\bibnamefont {Wang}}, \bibinfo {author} {\bibfnamefont {W.}~\bibnamefont {Yi}},\ and\ \bibinfo {author} {\bibfnamefont {P.}~\bibnamefont {Xue}},\ }\bibfield  {title} {\bibinfo {title} {Non-hermitian bulk–boundary correspondence in quantum dynamics},\ }\href {https://doi.org/10.1038/s41567-020-0836-6} {\bibfield  {journal} {\bibinfo  {journal} {Nature Physics}\ }\textbf {\bibinfo {volume} {16}},\ \bibinfo {pages} {761–766} (\bibinfo {year} {2020})}\BibitemShut {NoStop}%
\bibitem [{\citenamefont {Wang}\ \emph {et~al.}(2021{\natexlab{a}})\citenamefont {Wang}, \citenamefont {Dutt}, \citenamefont {Yang}, \citenamefont {Wojcik}, \citenamefont {Vučković},\ and\ \citenamefont {Fan}}]{wang_generating_2021}%
  \BibitemOpen
  \bibfield  {author} {\bibinfo {author} {\bibfnamefont {K.}~\bibnamefont {Wang}}, \bibinfo {author} {\bibfnamefont {A.}~\bibnamefont {Dutt}}, \bibinfo {author} {\bibfnamefont {K.~Y.}\ \bibnamefont {Yang}}, \bibinfo {author} {\bibfnamefont {C.~C.}\ \bibnamefont {Wojcik}}, \bibinfo {author} {\bibfnamefont {J.}~\bibnamefont {Vučković}},\ and\ \bibinfo {author} {\bibfnamefont {S.}~\bibnamefont {Fan}},\ }\bibfield  {title} {\bibinfo {title} {Generating arbitrary topological windings of a non-{Hermitian} band},\ }\href {https://doi.org/10.1126/science.abf6568} {\bibfield  {journal} {\bibinfo  {journal} {Science}\ }\textbf {\bibinfo {volume} {371}},\ \bibinfo {pages} {1240} (\bibinfo {year} {2021}{\natexlab{a}})}\BibitemShut {NoStop}%
\bibitem [{\citenamefont {Wang}\ \emph {et~al.}(2021{\natexlab{b}})\citenamefont {Wang}, \citenamefont {Dutt}, \citenamefont {Wojcik},\ and\ \citenamefont {Fan}}]{wang_topological_2021}%
  \BibitemOpen
  \bibfield  {author} {\bibinfo {author} {\bibfnamefont {K.}~\bibnamefont {Wang}}, \bibinfo {author} {\bibfnamefont {A.}~\bibnamefont {Dutt}}, \bibinfo {author} {\bibfnamefont {C.~C.}\ \bibnamefont {Wojcik}},\ and\ \bibinfo {author} {\bibfnamefont {S.}~\bibnamefont {Fan}},\ }\bibfield  {title} {\bibinfo {title} {Topological complex-energy braiding of non-{Hermitian} bands},\ }\href {https://doi.org/10.1038/s41586-021-03848-x} {\bibfield  {journal} {\bibinfo  {journal} {Nature}\ }\textbf {\bibinfo {volume} {598}},\ \bibinfo {pages} {59} (\bibinfo {year} {2021}{\natexlab{b}})}\BibitemShut {NoStop}%
\bibitem [{\citenamefont {Ye}\ \emph {et~al.}(2025{\natexlab{a}})\citenamefont {Ye}, \citenamefont {He}, \citenamefont {Li}, \citenamefont {Wang}, \citenamefont {Wu}, \citenamefont {Qiao}, \citenamefont {Zheng}, \citenamefont {Jin}, \citenamefont {Wang}, \citenamefont {Yuan},\ and\ \citenamefont {Chen}}]{Ye_observing_2025}%
  \BibitemOpen
  \bibfield  {author} {\bibinfo {author} {\bibfnamefont {R.}~\bibnamefont {Ye}}, \bibinfo {author} {\bibfnamefont {Y.}~\bibnamefont {He}}, \bibinfo {author} {\bibfnamefont {G.}~\bibnamefont {Li}}, \bibinfo {author} {\bibfnamefont {L.}~\bibnamefont {Wang}}, \bibinfo {author} {\bibfnamefont {X.}~\bibnamefont {Wu}}, \bibinfo {author} {\bibfnamefont {X.}~\bibnamefont {Qiao}}, \bibinfo {author} {\bibfnamefont {Y.}~\bibnamefont {Zheng}}, \bibinfo {author} {\bibfnamefont {L.}~\bibnamefont {Jin}}, \bibinfo {author} {\bibfnamefont {D.-W.}\ \bibnamefont {Wang}}, \bibinfo {author} {\bibfnamefont {L.}~\bibnamefont {Yuan}},\ and\ \bibinfo {author} {\bibfnamefont {X.}~\bibnamefont {Chen}},\ }\bibfield  {title} {\bibinfo {title} {Observing non-hermiticity induced chirality breaking in a synthetic hall ladder},\ }\bibfield  {journal} {\bibinfo  {journal} {Light: Science \& Applications}\ }\textbf {\bibinfo {volume} {14}},\ \href {https://doi.org/10.1038/s41377-024-01700-1} {10.1038/s41377-024-01700-1} (\bibinfo {year}
  {2025}{\natexlab{a}})\BibitemShut {NoStop}%
\bibitem [{\citenamefont {Örsel}\ \emph {et~al.}(2025)\citenamefont {Örsel}, \citenamefont {Noh}, \citenamefont {Zhu}, \citenamefont {Yim}, \citenamefont {Hughes}, \citenamefont {Thomale},\ and\ \citenamefont {Bahl}}]{orsel_giant_2025}%
  \BibitemOpen
  \bibfield  {author} {\bibinfo {author} {\bibfnamefont {O.~E.}\ \bibnamefont {Örsel}}, \bibinfo {author} {\bibfnamefont {J.}~\bibnamefont {Noh}}, \bibinfo {author} {\bibfnamefont {P.}~\bibnamefont {Zhu}}, \bibinfo {author} {\bibfnamefont {J.}~\bibnamefont {Yim}}, \bibinfo {author} {\bibfnamefont {T.~L.}\ \bibnamefont {Hughes}}, \bibinfo {author} {\bibfnamefont {R.}~\bibnamefont {Thomale}},\ and\ \bibinfo {author} {\bibfnamefont {G.}~\bibnamefont {Bahl}},\ }\bibfield  {title} {\bibinfo {title} {Giant {Nonreciprocity} and {Gyration} through {Modulation}-{Induced} {Hatano}-{Nelson} {Coupling} in {Integrated} {Photonics}},\ }\href {https://doi.org/10.1103/PhysRevLett.134.153801} {\bibfield  {journal} {\bibinfo  {journal} {Phys. Rev. Lett.}\ }\textbf {\bibinfo {volume} {134}},\ \bibinfo {pages} {153801} (\bibinfo {year} {2025})}\BibitemShut {NoStop}%
\bibitem [{\citenamefont {Blanchard}\ \emph {et~al.}(2025)\citenamefont {Blanchard}, \citenamefont {McDonald},\ and\ \citenamefont {St-Jean}}]{blanchard_exponentially_2025}%
  \BibitemOpen
  \bibfield  {author} {\bibinfo {author} {\bibfnamefont {P.-E.}\ \bibnamefont {Blanchard}}, \bibinfo {author} {\bibfnamefont {A.}~\bibnamefont {McDonald}},\ and\ \bibinfo {author} {\bibfnamefont {P.}~\bibnamefont {St-Jean}},\ }\href {https://doi.org/10.48550/arXiv.2511.16895} {\bibinfo {title} {Exponentially enhanced sensing through nonreciprocal light propagation}} (\bibinfo {year} {2025}),\ \bibinfo {note} {arXiv:2511.16895 [physics]}\BibitemShut {NoStop}%
\bibitem [{\citenamefont {Yu}\ \emph {et~al.}(2025)\citenamefont {Yu}, \citenamefont {Song}, \citenamefont {Wang}, \citenamefont {Srikanth}, \citenamefont {Sridhar}, \citenamefont {Chen}, \citenamefont {Huang}, \citenamefont {Li}, \citenamefont {Qiao}, \citenamefont {Wu}, \citenamefont {Dong}, \citenamefont {He}, \citenamefont {Xiao}, \citenamefont {Chen}, \citenamefont {Dutt}, \citenamefont {Gadway},\ and\ \citenamefont {Yuan}}]{yu_comprehensive_2025}%
  \BibitemOpen
  \bibfield  {author} {\bibinfo {author} {\bibfnamefont {D.}~\bibnamefont {Yu}}, \bibinfo {author} {\bibfnamefont {W.}~\bibnamefont {Song}}, \bibinfo {author} {\bibfnamefont {L.}~\bibnamefont {Wang}}, \bibinfo {author} {\bibfnamefont {R.}~\bibnamefont {Srikanth}}, \bibinfo {author} {\bibfnamefont {S.~K.}\ \bibnamefont {Sridhar}}, \bibinfo {author} {\bibfnamefont {T.}~\bibnamefont {Chen}}, \bibinfo {author} {\bibfnamefont {C.}~\bibnamefont {Huang}}, \bibinfo {author} {\bibfnamefont {G.}~\bibnamefont {Li}}, \bibinfo {author} {\bibfnamefont {X.}~\bibnamefont {Qiao}}, \bibinfo {author} {\bibfnamefont {X.}~\bibnamefont {Wu}}, \bibinfo {author} {\bibfnamefont {Z.}~\bibnamefont {Dong}}, \bibinfo {author} {\bibfnamefont {Y.}~\bibnamefont {He}}, \bibinfo {author} {\bibfnamefont {M.}~\bibnamefont {Xiao}}, \bibinfo {author} {\bibfnamefont {X.}~\bibnamefont {Chen}}, \bibinfo {author} {\bibfnamefont {A.}~\bibnamefont {Dutt}}, \bibinfo {author} {\bibfnamefont {B.}~\bibnamefont {Gadway}},\ and\ \bibinfo {author}
  {\bibfnamefont {L.}~\bibnamefont {Yuan}},\ }\bibfield  {title} {\bibinfo {title} {Comprehensive review on developments of synthetic dimensions},\ }\href {https://doi.org/10.3788/PI.2025.R06} {\bibfield  {journal} {\bibinfo  {journal} {pi}\ }\textbf {\bibinfo {volume} {4}},\ \bibinfo {pages} {R06} (\bibinfo {year} {2025})}\BibitemShut {NoStop}%
\bibitem [{\citenamefont {Koch}\ and\ \citenamefont {Budich}(2020)}]{koch_bulk-boundary_2020}%
  \BibitemOpen
  \bibfield  {author} {\bibinfo {author} {\bibfnamefont {R.}~\bibnamefont {Koch}}\ and\ \bibinfo {author} {\bibfnamefont {J.~C.}\ \bibnamefont {Budich}},\ }\bibfield  {title} {\bibinfo {title} {Bulk-boundary correspondence in non-{Hermitian} systems: stability analysis for generalized boundary conditions},\ }\href {https://doi.org/10.1140/epjd/e2020-100641-y} {\bibfield  {journal} {\bibinfo  {journal} {Eur. Phys. J. D}\ }\textbf {\bibinfo {volume} {74}},\ \bibinfo {pages} {70} (\bibinfo {year} {2020})}\BibitemShut {NoStop}%
\bibitem [{\citenamefont {Wang}\ \emph {et~al.}(2024)\citenamefont {Wang}, \citenamefont {Zhong},\ and\ \citenamefont {Fan}}]{Wang_non-hermitian_2024}%
  \BibitemOpen
  \bibfield  {author} {\bibinfo {author} {\bibfnamefont {H.}~\bibnamefont {Wang}}, \bibinfo {author} {\bibfnamefont {J.}~\bibnamefont {Zhong}},\ and\ \bibinfo {author} {\bibfnamefont {S.}~\bibnamefont {Fan}},\ }\bibfield  {title} {\bibinfo {title} {Non-hermitian photonic band winding and skin effects: a tutorial},\ }\href {https://doi.org/10.1364/aop.529289} {\bibfield  {journal} {\bibinfo  {journal} {Advances in Optics and Photonics}\ }\textbf {\bibinfo {volume} {16}},\ \bibinfo {pages} {659} (\bibinfo {year} {2024})}\BibitemShut {NoStop}%
\bibitem [{\citenamefont {Sridhar}\ \emph {et~al.}(2025)\citenamefont {Sridhar}, \citenamefont {Srikanth}, \citenamefont {Miller}, \citenamefont {McComb},\ and\ \citenamefont {Dutt}}]{sridhar_measuring_2025}%
  \BibitemOpen
  \bibfield  {author} {\bibinfo {author} {\bibfnamefont {S.~K.}\ \bibnamefont {Sridhar}}, \bibinfo {author} {\bibfnamefont {R.}~\bibnamefont {Srikanth}}, \bibinfo {author} {\bibfnamefont {A.~R.}\ \bibnamefont {Miller}}, \bibinfo {author} {\bibfnamefont {F.~J.}\ \bibnamefont {McComb}},\ and\ \bibinfo {author} {\bibfnamefont {A.}~\bibnamefont {Dutt}},\ }\href {https://doi.org/10.48550/arXiv.2505.04151} {\bibinfo {title} {Measuring {Z2} invariants in dimer models and cross-coupled ladders with a programmable photonic molecule}} (\bibinfo {year} {2025})\BibitemShut {NoStop}%
\bibitem [{\citenamefont {Dutt}\ \emph {et~al.}(2022)\citenamefont {Dutt}, \citenamefont {Yuan}, \citenamefont {Yang}, \citenamefont {Wang}, \citenamefont {Buddhiraju}, \citenamefont {Vučković},\ and\ \citenamefont {Fan}}]{dutt_creating_2022}%
  \BibitemOpen
  \bibfield  {author} {\bibinfo {author} {\bibfnamefont {A.}~\bibnamefont {Dutt}}, \bibinfo {author} {\bibfnamefont {L.}~\bibnamefont {Yuan}}, \bibinfo {author} {\bibfnamefont {K.~Y.}\ \bibnamefont {Yang}}, \bibinfo {author} {\bibfnamefont {K.}~\bibnamefont {Wang}}, \bibinfo {author} {\bibfnamefont {S.}~\bibnamefont {Buddhiraju}}, \bibinfo {author} {\bibfnamefont {J.}~\bibnamefont {Vučković}},\ and\ \bibinfo {author} {\bibfnamefont {S.}~\bibnamefont {Fan}},\ }\bibfield  {title} {\bibinfo {title} {Creating boundaries along a synthetic frequency dimension},\ }\href {https://doi.org/10.1038/s41467-022-31140-7} {\bibfield  {journal} {\bibinfo  {journal} {Nat Commun}\ }\textbf {\bibinfo {volume} {13}},\ \bibinfo {pages} {3377} (\bibinfo {year} {2022})}\BibitemShut {NoStop}%
\bibitem [{\citenamefont {Hu}\ \emph {et~al.}(2022)\citenamefont {Hu}, \citenamefont {Yu}, \citenamefont {Sinclair}, \citenamefont {Zhu}, \citenamefont {Cheng}, \citenamefont {Wang},\ and\ \citenamefont {Lončar}}]{hu_mirror-induced_2022}%
  \BibitemOpen
  \bibfield  {author} {\bibinfo {author} {\bibfnamefont {Y.}~\bibnamefont {Hu}}, \bibinfo {author} {\bibfnamefont {M.}~\bibnamefont {Yu}}, \bibinfo {author} {\bibfnamefont {N.}~\bibnamefont {Sinclair}}, \bibinfo {author} {\bibfnamefont {D.}~\bibnamefont {Zhu}}, \bibinfo {author} {\bibfnamefont {R.}~\bibnamefont {Cheng}}, \bibinfo {author} {\bibfnamefont {C.}~\bibnamefont {Wang}},\ and\ \bibinfo {author} {\bibfnamefont {M.}~\bibnamefont {Lončar}},\ }\bibfield  {title} {\bibinfo {title} {Mirror-induced reflection in the frequency domain},\ }\href {https://doi.org/10.1038/s41467-022-33529-w} {\bibfield  {journal} {\bibinfo  {journal} {Nat Commun}\ }\textbf {\bibinfo {volume} {13}},\ \bibinfo {pages} {6293} (\bibinfo {year} {2022})},\ \bibinfo {note} {number: 1}\BibitemShut {NoStop}%
\bibitem [{\citenamefont {Dutt}\ \emph {et~al.}(2019)\citenamefont {Dutt}, \citenamefont {Minkov}, \citenamefont {Lin}, \citenamefont {Yuan}, \citenamefont {Miller},\ and\ \citenamefont {Fan}}]{dutt_experimental_2019-1}%
  \BibitemOpen
  \bibfield  {author} {\bibinfo {author} {\bibfnamefont {A.}~\bibnamefont {Dutt}}, \bibinfo {author} {\bibfnamefont {M.}~\bibnamefont {Minkov}}, \bibinfo {author} {\bibfnamefont {Q.}~\bibnamefont {Lin}}, \bibinfo {author} {\bibfnamefont {L.}~\bibnamefont {Yuan}}, \bibinfo {author} {\bibfnamefont {D.~A.~B.}\ \bibnamefont {Miller}},\ and\ \bibinfo {author} {\bibfnamefont {S.}~\bibnamefont {Fan}},\ }\bibfield  {title} {\bibinfo {title} {Experimental band structure spectroscopy along a synthetic dimension},\ }\href {https://doi.org/10.1038/s41467-019-11117-9} {\bibfield  {journal} {\bibinfo  {journal} {Nature Communications}\ }\textbf {\bibinfo {volume} {10}},\ \bibinfo {pages} {3122} (\bibinfo {year} {2019})}\BibitemShut {NoStop}%
\bibitem [{\citenamefont {Dutt}\ \emph {et~al.}(2020)\citenamefont {Dutt}, \citenamefont {Lin}, \citenamefont {Yuan}, \citenamefont {Minkov}, \citenamefont {Xiao},\ and\ \citenamefont {Fan}}]{dutt_single_2020}%
  \BibitemOpen
  \bibfield  {author} {\bibinfo {author} {\bibfnamefont {A.}~\bibnamefont {Dutt}}, \bibinfo {author} {\bibfnamefont {Q.}~\bibnamefont {Lin}}, \bibinfo {author} {\bibfnamefont {L.}~\bibnamefont {Yuan}}, \bibinfo {author} {\bibfnamefont {M.}~\bibnamefont {Minkov}}, \bibinfo {author} {\bibfnamefont {M.}~\bibnamefont {Xiao}},\ and\ \bibinfo {author} {\bibfnamefont {S.}~\bibnamefont {Fan}},\ }\bibfield  {title} {\bibinfo {title} {A single photonic cavity with two independent physical synthetic dimensions},\ }\href {https://doi.org/10.1126/science.aaz3071} {\bibfield  {journal} {\bibinfo  {journal} {Science}\ }\textbf {\bibinfo {volume} {367}},\ \bibinfo {pages} {59} (\bibinfo {year} {2020})}\BibitemShut {NoStop}%
\bibitem [{\citenamefont {Li}\ \emph {et~al.}(2021)\citenamefont {Li}, \citenamefont {Zheng}, \citenamefont {Dutt}, \citenamefont {Yu}, \citenamefont {Shan}, \citenamefont {Liu}, \citenamefont {Yuan}, \citenamefont {Fan},\ and\ \citenamefont {Chen}}]{li_dynamic_2021}%
  \BibitemOpen
  \bibfield  {author} {\bibinfo {author} {\bibfnamefont {G.}~\bibnamefont {Li}}, \bibinfo {author} {\bibfnamefont {Y.}~\bibnamefont {Zheng}}, \bibinfo {author} {\bibfnamefont {A.}~\bibnamefont {Dutt}}, \bibinfo {author} {\bibfnamefont {D.}~\bibnamefont {Yu}}, \bibinfo {author} {\bibfnamefont {Q.}~\bibnamefont {Shan}}, \bibinfo {author} {\bibfnamefont {S.}~\bibnamefont {Liu}}, \bibinfo {author} {\bibfnamefont {L.}~\bibnamefont {Yuan}}, \bibinfo {author} {\bibfnamefont {S.}~\bibnamefont {Fan}},\ and\ \bibinfo {author} {\bibfnamefont {X.}~\bibnamefont {Chen}},\ }\bibfield  {title} {\bibinfo {title} {Dynamic band structure measurement in the synthetic space},\ }\href {https://doi.org/10.1126/sciadv.abe4335} {\bibfield  {journal} {\bibinfo  {journal} {Science Advances}\ }\textbf {\bibinfo {volume} {7}},\ \bibinfo {pages} {eabe4335} (\bibinfo {year} {2021})}\BibitemShut {NoStop}%
\bibitem [{\citenamefont {Balčytis}\ \emph {et~al.}(2022)\citenamefont {Balčytis}, \citenamefont {Ozawa}, \citenamefont {Ota}, \citenamefont {Iwamoto}, \citenamefont {Maeda},\ and\ \citenamefont {Baba}}]{balcytis_synthetic_2022}%
  \BibitemOpen
  \bibfield  {author} {\bibinfo {author} {\bibfnamefont {A.}~\bibnamefont {Balčytis}}, \bibinfo {author} {\bibfnamefont {T.}~\bibnamefont {Ozawa}}, \bibinfo {author} {\bibfnamefont {Y.}~\bibnamefont {Ota}}, \bibinfo {author} {\bibfnamefont {S.}~\bibnamefont {Iwamoto}}, \bibinfo {author} {\bibfnamefont {J.}~\bibnamefont {Maeda}},\ and\ \bibinfo {author} {\bibfnamefont {T.}~\bibnamefont {Baba}},\ }\bibfield  {title} {\bibinfo {title} {Synthetic dimension band structures on a {Si} {CMOS} photonic platform},\ }\href {https://doi.org/10.1126/sciadv.abk0468} {\bibfield  {journal} {\bibinfo  {journal} {Science Advances}\ }\textbf {\bibinfo {volume} {8}},\ \bibinfo {pages} {eabk0468} (\bibinfo {year} {2022})}\BibitemShut {NoStop}%
\bibitem [{\citenamefont {Senanian}\ \emph {et~al.}(2023)\citenamefont {Senanian}, \citenamefont {Wright}, \citenamefont {Wade}, \citenamefont {Doyle},\ and\ \citenamefont {McMahon}}]{senanian_programmable_2023}%
  \BibitemOpen
  \bibfield  {author} {\bibinfo {author} {\bibfnamefont {A.}~\bibnamefont {Senanian}}, \bibinfo {author} {\bibfnamefont {L.~G.}\ \bibnamefont {Wright}}, \bibinfo {author} {\bibfnamefont {P.~F.}\ \bibnamefont {Wade}}, \bibinfo {author} {\bibfnamefont {H.~K.}\ \bibnamefont {Doyle}},\ and\ \bibinfo {author} {\bibfnamefont {P.~L.}\ \bibnamefont {McMahon}},\ }\bibfield  {title} {\bibinfo {title} {Programmable large-scale simulation of bosonic transport in optical synthetic frequency lattices},\ }\href {https://doi.org/10.1038/s41567-023-02075-7} {\bibfield  {journal} {\bibinfo  {journal} {Nat. Phys.}\ }\textbf {\bibinfo {volume} {19}},\ \bibinfo {pages} {1333} (\bibinfo {year} {2023})}\BibitemShut {NoStop}%
\bibitem [{\citenamefont {Li}\ \emph {et~al.}(2023)\citenamefont {Li}, \citenamefont {Wang}, \citenamefont {Ye}, \citenamefont {Zheng}, \citenamefont {Wang}, \citenamefont {Liu}, \citenamefont {Dutt}, \citenamefont {Yuan},\ and\ \citenamefont {Chen}}]{li_direct_2023}%
  \BibitemOpen
  \bibfield  {author} {\bibinfo {author} {\bibfnamefont {G.}~\bibnamefont {Li}}, \bibinfo {author} {\bibfnamefont {L.}~\bibnamefont {Wang}}, \bibinfo {author} {\bibfnamefont {R.}~\bibnamefont {Ye}}, \bibinfo {author} {\bibfnamefont {Y.}~\bibnamefont {Zheng}}, \bibinfo {author} {\bibfnamefont {D.-W.}\ \bibnamefont {Wang}}, \bibinfo {author} {\bibfnamefont {X.-J.}\ \bibnamefont {Liu}}, \bibinfo {author} {\bibfnamefont {A.}~\bibnamefont {Dutt}}, \bibinfo {author} {\bibfnamefont {L.}~\bibnamefont {Yuan}},\ and\ \bibinfo {author} {\bibfnamefont {X.}~\bibnamefont {Chen}},\ }\bibfield  {title} {\bibinfo {title} {Direct extraction of topological {Zak} phase with the synthetic dimension},\ }\href {https://doi.org/10.1038/s41377-023-01126-1} {\bibfield  {journal} {\bibinfo  {journal} {Light Sci Appl}\ }\textbf {\bibinfo {volume} {12}},\ \bibinfo {pages} {81} (\bibinfo {year} {2023})}\BibitemShut {NoStop}%
\bibitem [{\citenamefont {Cheng}\ \emph {et~al.}(2023)\citenamefont {Cheng}, \citenamefont {Lustig}, \citenamefont {Wang},\ and\ \citenamefont {Fan}}]{cheng_multidimensional_2023}%
  \BibitemOpen
  \bibfield  {author} {\bibinfo {author} {\bibfnamefont {D.}~\bibnamefont {Cheng}}, \bibinfo {author} {\bibfnamefont {E.}~\bibnamefont {Lustig}}, \bibinfo {author} {\bibfnamefont {K.}~\bibnamefont {Wang}},\ and\ \bibinfo {author} {\bibfnamefont {S.}~\bibnamefont {Fan}},\ }\bibfield  {title} {\bibinfo {title} {Multi-dimensional band structure spectroscopy in the synthetic frequency dimension},\ }\href {https://doi.org/10.1038/s41377-023-01196-1} {\bibfield  {journal} {\bibinfo  {journal} {Light: Science {\&} Applications}\ }\textbf {\bibinfo {volume} {12}},\ \bibinfo {pages} {158} (\bibinfo {year} {2023})}\BibitemShut {NoStop}%
\bibitem [{\citenamefont {Dinh}\ \emph {et~al.}(2024)\citenamefont {Dinh}, \citenamefont {Balčytis}, \citenamefont {Ozawa}, \citenamefont {Ota}, \citenamefont {Ren}, \citenamefont {Baba}, \citenamefont {Iwamoto}, \citenamefont {Mitchell},\ and\ \citenamefont {Nguyen}}]{dinh_reconfigurable_2024}%
  \BibitemOpen
  \bibfield  {author} {\bibinfo {author} {\bibfnamefont {H.~X.}\ \bibnamefont {Dinh}}, \bibinfo {author} {\bibfnamefont {A.}~\bibnamefont {Balčytis}}, \bibinfo {author} {\bibfnamefont {T.}~\bibnamefont {Ozawa}}, \bibinfo {author} {\bibfnamefont {Y.}~\bibnamefont {Ota}}, \bibinfo {author} {\bibfnamefont {G.}~\bibnamefont {Ren}}, \bibinfo {author} {\bibfnamefont {T.}~\bibnamefont {Baba}}, \bibinfo {author} {\bibfnamefont {S.}~\bibnamefont {Iwamoto}}, \bibinfo {author} {\bibfnamefont {A.}~\bibnamefont {Mitchell}},\ and\ \bibinfo {author} {\bibfnamefont {T.~G.}\ \bibnamefont {Nguyen}},\ }\bibfield  {title} {\bibinfo {title} {Reconfigurable synthetic dimension frequency lattices in an integrated lithium niobate ring cavity},\ }\href {https://doi.org/10.1038/s42005-024-01676-9} {\bibfield  {journal} {\bibinfo  {journal} {Commun Phys}\ }\textbf {\bibinfo {volume} {7}},\ \bibinfo {pages} {1} (\bibinfo {year} {2024})}\BibitemShut {NoStop}%
\bibitem [{\citenamefont {Pellerin}\ \emph {et~al.}(2024)\citenamefont {Pellerin}, \citenamefont {Houvenaghel}, \citenamefont {Coish}, \citenamefont {Carusotto},\ and\ \citenamefont {St-Jean}}]{pellerin_wave-function_2024}%
  \BibitemOpen
  \bibfield  {author} {\bibinfo {author} {\bibfnamefont {F.}~\bibnamefont {Pellerin}}, \bibinfo {author} {\bibfnamefont {R.}~\bibnamefont {Houvenaghel}}, \bibinfo {author} {\bibfnamefont {W.}~\bibnamefont {Coish}}, \bibinfo {author} {\bibfnamefont {I.}~\bibnamefont {Carusotto}},\ and\ \bibinfo {author} {\bibfnamefont {P.}~\bibnamefont {St-Jean}},\ }\bibfield  {title} {\bibinfo {title} {Wave-{Function} {Tomography} of {Topological} {Dimer} {Chains} with {Long}-{Range} {Couplings}},\ }\href {https://doi.org/10.1103/PhysRevLett.132.183802} {\bibfield  {journal} {\bibinfo  {journal} {Phys. Rev. Lett.}\ }\textbf {\bibinfo {volume} {132}},\ \bibinfo {pages} {183802} (\bibinfo {year} {2024})}\BibitemShut {NoStop}%
\bibitem [{\citenamefont {Ch\`enier}\ \emph {et~al.}(2024)\citenamefont {Ch\`enier}, \citenamefont {d'Aligny}, \citenamefont {Pellerin}, \citenamefont {Blanchard}, \citenamefont {Ozawa}, \citenamefont {Carusotto},\ and\ \citenamefont {St-Jean}}]{chenier_quantized_2024}%
  \BibitemOpen
  \bibfield  {author} {\bibinfo {author} {\bibfnamefont {A.}~\bibnamefont {Ch\`enier}}, \bibinfo {author} {\bibfnamefont {B.}~\bibnamefont {d'Aligny}}, \bibinfo {author} {\bibfnamefont {F.}~\bibnamefont {Pellerin}}, \bibinfo {author} {\bibfnamefont {P.-E.}\ \bibnamefont {Blanchard}}, \bibinfo {author} {\bibfnamefont {T.}~\bibnamefont {Ozawa}}, \bibinfo {author} {\bibfnamefont {I.}~\bibnamefont {Carusotto}},\ and\ \bibinfo {author} {\bibfnamefont {P.}~\bibnamefont {St-Jean}},\ }\bibfield  {title} {\bibinfo {title} {Quantized {Hall} drift in a frequency-encoded photonic {Chern} insulator},\ }\bibfield  {journal} {\bibinfo  {journal} {arXiv:2412.04347}\ }\href {https://doi.org/10.48550/arXiv.2412.04347} {10.48550/arXiv.2412.04347} (\bibinfo {year} {2024})\BibitemShut {NoStop}%
\bibitem [{\citenamefont {Wang}\ \emph {et~al.}(2025{\natexlab{b}})\citenamefont {Wang}, \citenamefont {Zeng}, \citenamefont {Wang}, \citenamefont {Ren}, \citenamefont {Ao}, \citenamefont {Li}, \citenamefont {Liu}, \citenamefont {Guo}, \citenamefont {Xie}, \citenamefont {Liu}, \citenamefont {Ma}, \citenamefont {Wu}, \citenamefont {Luo}, \citenamefont {Wang}, \citenamefont {Tang}, \citenamefont {Li},\ and\ \citenamefont {Guo}}]{Wang_versatile_2025}%
  \BibitemOpen
  \bibfield  {author} {\bibinfo {author} {\bibfnamefont {Z.-A.}\ \bibnamefont {Wang}}, \bibinfo {author} {\bibfnamefont {X.-D.}\ \bibnamefont {Zeng}}, \bibinfo {author} {\bibfnamefont {Y.-T.}\ \bibnamefont {Wang}}, \bibinfo {author} {\bibfnamefont {J.-M.}\ \bibnamefont {Ren}}, \bibinfo {author} {\bibfnamefont {C.}~\bibnamefont {Ao}}, \bibinfo {author} {\bibfnamefont {Z.-P.}\ \bibnamefont {Li}}, \bibinfo {author} {\bibfnamefont {W.}~\bibnamefont {Liu}}, \bibinfo {author} {\bibfnamefont {N.-J.}\ \bibnamefont {Guo}}, \bibinfo {author} {\bibfnamefont {L.-K.}\ \bibnamefont {Xie}}, \bibinfo {author} {\bibfnamefont {J.-Y.}\ \bibnamefont {Liu}}, \bibinfo {author} {\bibfnamefont {Y.-H.}\ \bibnamefont {Ma}}, \bibinfo {author} {\bibfnamefont {Y.-Q.}\ \bibnamefont {Wu}}, \bibinfo {author} {\bibfnamefont {X.-W.}\ \bibnamefont {Luo}}, \bibinfo {author} {\bibfnamefont {S.}~\bibnamefont {Wang}}, \bibinfo {author} {\bibfnamefont {J.-S.}\ \bibnamefont {Tang}}, \bibinfo {author} {\bibfnamefont {C.-F.}\ \bibnamefont {Li}},\ and\
  \bibinfo {author} {\bibfnamefont {G.-C.}\ \bibnamefont {Guo}},\ }\bibfield  {title} {\bibinfo {title} {Versatile photonic frequency synthetic dimensions using a single programmable on-chip device},\ }\bibfield  {journal} {\bibinfo  {journal} {Nature Communications}\ }\textbf {\bibinfo {volume} {16}},\ \href {https://doi.org/10.1038/s41467-025-63114-w} {10.1038/s41467-025-63114-w} (\bibinfo {year} {2025}{\natexlab{b}})\BibitemShut {NoStop}%
\bibitem [{\citenamefont {Ye}\ \emph {et~al.}(2025{\natexlab{b}})\citenamefont {Ye}, \citenamefont {Li}, \citenamefont {Wan}, \citenamefont {Xue}, \citenamefont {Wang}, \citenamefont {Qiao}, \citenamefont {Wang}, \citenamefont {Li}, \citenamefont {Liu}, \citenamefont {Wang}, \citenamefont {Ma}, \citenamefont {Bo}, \citenamefont {Zheng}, \citenamefont {Dong}, \citenamefont {Yuan},\ and\ \citenamefont {Chen}}]{ye_construction_2025}%
  \BibitemOpen
  \bibfield  {author} {\bibinfo {author} {\bibfnamefont {R.}~\bibnamefont {Ye}}, \bibinfo {author} {\bibfnamefont {G.}~\bibnamefont {Li}}, \bibinfo {author} {\bibfnamefont {S.}~\bibnamefont {Wan}}, \bibinfo {author} {\bibfnamefont {X.}~\bibnamefont {Xue}}, \bibinfo {author} {\bibfnamefont {P.-Y.}\ \bibnamefont {Wang}}, \bibinfo {author} {\bibfnamefont {X.}~\bibnamefont {Qiao}}, \bibinfo {author} {\bibfnamefont {L.}~\bibnamefont {Wang}}, \bibinfo {author} {\bibfnamefont {H.}~\bibnamefont {Li}}, \bibinfo {author} {\bibfnamefont {S.}~\bibnamefont {Liu}}, \bibinfo {author} {\bibfnamefont {J.}~\bibnamefont {Wang}}, \bibinfo {author} {\bibfnamefont {R.}~\bibnamefont {Ma}}, \bibinfo {author} {\bibfnamefont {F.}~\bibnamefont {Bo}}, \bibinfo {author} {\bibfnamefont {Y.}~\bibnamefont {Zheng}}, \bibinfo {author} {\bibfnamefont {C.-H.}\ \bibnamefont {Dong}}, \bibinfo {author} {\bibfnamefont {L.}~\bibnamefont {Yuan}},\ and\ \bibinfo {author} {\bibfnamefont {X.}~\bibnamefont {Chen}},\ }\bibfield  {title} {\bibinfo {title}
  {Construction of various time-varying hamiltonians on thin-film lithium niobate chip},\ }\href {https://doi.org/10.1103/PhysRevLett.134.163802} {\bibfield  {journal} {\bibinfo  {journal} {Phys. Rev. Lett.}\ }\textbf {\bibinfo {volume} {134}},\ \bibinfo {pages} {163802} (\bibinfo {year} {2025}{\natexlab{b}})}\BibitemShut {NoStop}%
\bibitem [{\citenamefont {Zhang}\ \emph {et~al.}(2019)\citenamefont {Zhang}, \citenamefont {Buscaino}, \citenamefont {Wang}, \citenamefont {Shams-Ansari}, \citenamefont {Reimer}, \citenamefont {Zhu}, \citenamefont {Kahn},\ and\ \citenamefont {Lončar}}]{zhang_broadband_2019}%
  \BibitemOpen
  \bibfield  {author} {\bibinfo {author} {\bibfnamefont {M.}~\bibnamefont {Zhang}}, \bibinfo {author} {\bibfnamefont {B.}~\bibnamefont {Buscaino}}, \bibinfo {author} {\bibfnamefont {C.}~\bibnamefont {Wang}}, \bibinfo {author} {\bibfnamefont {A.}~\bibnamefont {Shams-Ansari}}, \bibinfo {author} {\bibfnamefont {C.}~\bibnamefont {Reimer}}, \bibinfo {author} {\bibfnamefont {R.}~\bibnamefont {Zhu}}, \bibinfo {author} {\bibfnamefont {J.~M.}\ \bibnamefont {Kahn}},\ and\ \bibinfo {author} {\bibfnamefont {M.}~\bibnamefont {Lončar}},\ }\bibfield  {title} {\bibinfo {title} {Broadband electro-optic frequency comb generation in a lithium niobate microring resonator},\ }\href {https://doi.org/10.1038/s41586-019-1008-7} {\bibfield  {journal} {\bibinfo  {journal} {Nature}\ }\textbf {\bibinfo {volume} {568}},\ \bibinfo {pages} {373} (\bibinfo {year} {2019})}\BibitemShut {NoStop}%
\bibitem [{\citenamefont {Yang}\ \emph {et~al.}(2025)\citenamefont {Yang}, \citenamefont {Fang}, \citenamefont {Zhang},\ and\ \citenamefont {Fang}}]{Yang_tailoring_2025}%
  \BibitemOpen
  \bibfield  {author} {\bibinfo {author} {\bibfnamefont {A.}~\bibnamefont {Yang}}, \bibinfo {author} {\bibfnamefont {Z.}~\bibnamefont {Fang}}, \bibinfo {author} {\bibfnamefont {K.}~\bibnamefont {Zhang}},\ and\ \bibinfo {author} {\bibfnamefont {C.}~\bibnamefont {Fang}},\ }\bibfield  {title} {\bibinfo {title} {Tailoring bound state geometry in high-dimensional non-hermitian systems},\ }\bibfield  {journal} {\bibinfo  {journal} {Communications Physics}\ }\textbf {\bibinfo {volume} {8}},\ \href {https://doi.org/10.1038/s42005-025-02037-w} {10.1038/s42005-025-02037-w} (\bibinfo {year} {2025})\BibitemShut {NoStop}%
\bibitem [{\citenamefont {Montag}\ and\ \citenamefont {Kunst}(2024)}]{Montag_symmetry-induced_2024}%
  \BibitemOpen
  \bibfield  {author} {\bibinfo {author} {\bibfnamefont {A.}~\bibnamefont {Montag}}\ and\ \bibinfo {author} {\bibfnamefont {F.~K.}\ \bibnamefont {Kunst}},\ }\bibfield  {title} {\bibinfo {title} {Symmetry-induced higher-order exceptional points in two dimensions},\ }\href {https://doi.org/10.1103/PhysRevResearch.6.023205} {\bibfield  {journal} {\bibinfo  {journal} {Phys. Rev. Res.}\ }\textbf {\bibinfo {volume} {6}},\ \bibinfo {pages} {023205} (\bibinfo {year} {2024})}\BibitemShut {NoStop}%
\bibitem [{\citenamefont {Zhao}\ \emph {et~al.}(2019)\citenamefont {Zhao}, \citenamefont {Qiao}, \citenamefont {Wu}, \citenamefont {Midya}, \citenamefont {Longhi},\ and\ \citenamefont {Feng}}]{zhao_non-hermitian_2019}%
  \BibitemOpen
  \bibfield  {author} {\bibinfo {author} {\bibfnamefont {H.}~\bibnamefont {Zhao}}, \bibinfo {author} {\bibfnamefont {X.}~\bibnamefont {Qiao}}, \bibinfo {author} {\bibfnamefont {T.}~\bibnamefont {Wu}}, \bibinfo {author} {\bibfnamefont {B.}~\bibnamefont {Midya}}, \bibinfo {author} {\bibfnamefont {S.}~\bibnamefont {Longhi}},\ and\ \bibinfo {author} {\bibfnamefont {L.}~\bibnamefont {Feng}},\ }\bibfield  {title} {\bibinfo {title} {Non-{Hermitian} topological light steering},\ }\href {https://doi.org/10.1126/science.aay1064} {\bibfield  {journal} {\bibinfo  {journal} {Science}\ }\textbf {\bibinfo {volume} {365}},\ \bibinfo {pages} {1163} (\bibinfo {year} {2019})}\BibitemShut {NoStop}%
\bibitem [{\citenamefont {Zou}\ \emph {et~al.}(2021)\citenamefont {Zou}, \citenamefont {Chen}, \citenamefont {He}, \citenamefont {Bao}, \citenamefont {Lee}, \citenamefont {Sun},\ and\ \citenamefont {Zhang}}]{Zou_observation_2021}%
  \BibitemOpen
  \bibfield  {author} {\bibinfo {author} {\bibfnamefont {D.}~\bibnamefont {Zou}}, \bibinfo {author} {\bibfnamefont {T.}~\bibnamefont {Chen}}, \bibinfo {author} {\bibfnamefont {W.}~\bibnamefont {He}}, \bibinfo {author} {\bibfnamefont {J.}~\bibnamefont {Bao}}, \bibinfo {author} {\bibfnamefont {C.~H.}\ \bibnamefont {Lee}}, \bibinfo {author} {\bibfnamefont {H.}~\bibnamefont {Sun}},\ and\ \bibinfo {author} {\bibfnamefont {X.}~\bibnamefont {Zhang}},\ }\bibfield  {title} {\bibinfo {title} {Observation of hybrid higher-order skin-topological effect in non-hermitian topolectrical circuits},\ }\bibfield  {journal} {\bibinfo  {journal} {Nature Communications}\ }\textbf {\bibinfo {volume} {12}},\ \href {https://doi.org/10.1038/s41467-021-26414-5} {10.1038/s41467-021-26414-5} (\bibinfo {year} {2021})\BibitemShut {NoStop}%
\bibitem [{\citenamefont {Zheng}\ \emph {et~al.}(2024)\citenamefont {Zheng}, \citenamefont {Jalali~Mehrabad}, \citenamefont {Vannucci}, \citenamefont {Li}, \citenamefont {Dutt}, \citenamefont {Hafezi}, \citenamefont {Mittal},\ and\ \citenamefont {Waks}}]{zheng_dynamic_2024}%
  \BibitemOpen
  \bibfield  {author} {\bibinfo {author} {\bibfnamefont {X.}~\bibnamefont {Zheng}}, \bibinfo {author} {\bibfnamefont {M.}~\bibnamefont {Jalali~Mehrabad}}, \bibinfo {author} {\bibfnamefont {J.}~\bibnamefont {Vannucci}}, \bibinfo {author} {\bibfnamefont {K.}~\bibnamefont {Li}}, \bibinfo {author} {\bibfnamefont {A.}~\bibnamefont {Dutt}}, \bibinfo {author} {\bibfnamefont {M.}~\bibnamefont {Hafezi}}, \bibinfo {author} {\bibfnamefont {S.}~\bibnamefont {Mittal}},\ and\ \bibinfo {author} {\bibfnamefont {E.}~\bibnamefont {Waks}},\ }\bibfield  {title} {\bibinfo {title} {Dynamic control of {2D} non-{Hermitian} photonic corner skin modes in synthetic dimensions},\ }\href {https://doi.org/10.1038/s41467-024-55236-4} {\bibfield  {journal} {\bibinfo  {journal} {Nat Commun}\ }\textbf {\bibinfo {volume} {15}},\ \bibinfo {pages} {10881} (\bibinfo {year} {2024})}\BibitemShut {NoStop}%
\bibitem [{\citenamefont {Yuan}\ \emph {et~al.}(2018)\citenamefont {Yuan}, \citenamefont {Xiao}, \citenamefont {Lin},\ and\ \citenamefont {Fan}}]{yuan_synthetic-space_2018}%
  \BibitemOpen
  \bibfield  {author} {\bibinfo {author} {\bibfnamefont {L.}~\bibnamefont {Yuan}}, \bibinfo {author} {\bibfnamefont {M.}~\bibnamefont {Xiao}}, \bibinfo {author} {\bibfnamefont {Q.}~\bibnamefont {Lin}},\ and\ \bibinfo {author} {\bibfnamefont {S.}~\bibnamefont {Fan}},\ }\bibfield  {title} {\bibinfo {title} {Synthetic-space with arbitrary dimensions in a few rings undergoing dynamic modulation},\ }\href {https://doi.org/10.1103/PhysRevB.97.104105} {\bibfield  {journal} {\bibinfo  {journal} {Physical Review B}\ }\textbf {\bibinfo {volume} {97}},\ \bibinfo {pages} {104105} (\bibinfo {year} {2018})}\BibitemShut {NoStop}%
\bibitem [{\citenamefont {Wang}\ \emph {et~al.}(2020)\citenamefont {Wang}, \citenamefont {Bell}, \citenamefont {Solntsev}, \citenamefont {Neshev}, \citenamefont {Eggleton},\ and\ \citenamefont {Sukhorukov}}]{wang_multidimensional_2020}%
  \BibitemOpen
  \bibfield  {author} {\bibinfo {author} {\bibfnamefont {K.}~\bibnamefont {Wang}}, \bibinfo {author} {\bibfnamefont {B.~A.}\ \bibnamefont {Bell}}, \bibinfo {author} {\bibfnamefont {A.~S.}\ \bibnamefont {Solntsev}}, \bibinfo {author} {\bibfnamefont {D.~N.}\ \bibnamefont {Neshev}}, \bibinfo {author} {\bibfnamefont {B.~J.}\ \bibnamefont {Eggleton}},\ and\ \bibinfo {author} {\bibfnamefont {A.~A.}\ \bibnamefont {Sukhorukov}},\ }\bibfield  {title} {\bibinfo {title} {Multidimensional synthetic chiral-tube lattices via nonlinear frequency conversion},\ }\href {https://doi.org/10.1038/s41377-020-0299-7} {\bibfield  {journal} {\bibinfo  {journal} {Light: Science \& Applications}\ }\textbf {\bibinfo {volume} {9}},\ \bibinfo {pages} {132} (\bibinfo {year} {2020})}\BibitemShut {NoStop}%
\end{thebibliography}%

\end{document}